\documentclass[aps,pre,twocolumn,superscriptaddress,showpacs]{revtex4}
\usepackage{graphicx}

\def\Z{{\bf Z}}
\def\R{{\bf R}}
\def\C{{\bf C}}
\def\x{{\bf x}}
\def\y{{\bf y}}
\def\z{{\bf z}}
\def\a{{\bf a}}
\def\b{{\bf b}}
\def\0{{\bf 0}}
\def\h{\widehat{h}_N}
\def\f{\widehat{f}_N}
\def\hN{\widehat{\cal N}_N}
\def\gT{\widehat{g}_{N,T}}
\def\p{\widehat{p}_N}

\def\rR{{\rm R}}
\def\rI{{\rm I}}

\begin{document}

\title{Vicious walk with a wall, noncolliding meanders,\\
and chiral and Bogoliubov-deGennes random matrices}

\author{Makoto Katori}
\email[]{katori@phys.chuo-u.ac.jp}
\affiliation{Department of Physics,
Faculty of Science and Engineering,
Chuo University, Kasuga, Bunkyo-ku, Tokyo 112-8551, Japan}
\author{Hideki Tanemura}
\email[]{tanemura@math.s.chiba-u.ac.jp}
\affiliation{
Department of Mathematics and Informatics,
Faculty of Science, \\
Chiba University, 1-33 Yayoi-cho, Inage-ku,
Chiba 263-8522, Japan}
\author{Taro Nagao}
\email[]{nagao@sphinx.phys.sci.osaka-u.ac.jp}
\affiliation{
Department of Physics, 
Graduate School of Science, 
Osaka University, 
Toyonaka, Osaka 560-0043, Japan}
\author{Naoaki Komatsuda}
\email[]{komatuda@phys.chuo-u.ac.jp}
\affiliation{Department of Physics,
Faculty of Science and Engineering,
Chuo University, Kasuga, Bunkyo-ku, Tokyo 112-8551, Japan}

\date{June 23, 2003}

\begin{abstract}

Spatially and temporally inhomogeneous evolution of
one-dimensional vicious walkers with wall restriction is studied.
We show that its continuum version is equivalent with
a noncolliding system of stochastic processes
called Brownian meanders. 
Here the Brownian meander is a temporally inhomogeneous process
introduced by Yor as a transform of the Bessel process that is 
a motion of radial coordinate of the three-dimensional
Brownian motion represented in the spherical coordinates.
It is proved that the spatial distribution of vicious walkers
with a wall at the origin can be described by
the eigenvalue-statistics of Gaussian ensembles of
Bogoliubov-deGennes Hamiltonians of the mean-field
theory of superconductivity, which have the particle-hole
symmetry. We report that the time evolution of the present
stochastic process is fully characterized by the change of
symmetry classes from the type $C$ to the type $C$I
in the nonstandard classes of random
matrix theory of Altland and Zirnbauer.
The relation between the noncolliding systems of the
generalized meanders of Yor,
which are associated with the even-dimensional
Bessel processes, and the chiral random matrix theory
is also clarified.

\end{abstract}


\maketitle

\section{INTRODUCTION} 

Unbalance between short-ranged properties of interactions
among elements and long-ranged cooperative effects
realized in macroscopic levels is a significant feature
of systems far from equilibrium.
Even in one dimension the contact process, a model of
infection of a contagious disease, exhibits a continuous
phase transition at a critical value $\lambda_{c}$
of the infection rate $\lambda$ and in $\lambda > \lambda_{c}$
infected and healthy individuals establish coexistence
without detailed balance \cite{Lig85,KKon93}. Boundary conditions
locally imposed at the two edges in one-dimensional lattice
play essential role to determine the bulk properties
in the asymmetric simple exclusion process,
which can be regarded as a model of traffic flows in highways
\cite{Lig99,DEHP93,Sas99}.
The purpose of the present paper is to propose one theoretical
treatment of such emergence of long-range effects in
simple ({\it i.e.} short-ranged) stochastic models
using vicious walker models originally introduced by
Fisher for wetting and melting transitions \cite{Fis84}.
The key-point is the symmetry of higher-dimensional space,
in which the nonequilibrium many-body system is embedded.

Consider $N$ identical and independent simple and
symmetric random walks with initial positions 
$x_{1} < x_{2} < \cdots < x_{N}$,
where $x_{j}$ are assumed to be even integers.
One of the fundamental quantities in the vicious walk 
problem \cite{CK03} is the probability ${\cal N}_{N}(t,\x),
\x=(x_{1}, \cdots, x_{N})$,
that all walkers remain the ordering of their positions
up to time $t$;
$x_{1}(s) < x_{2}(s) < \cdots < x_{N}(s)$ for all
$0 \leq s \leq t$. 
In other words, it is the probability that
they never collide with each other for
a time period $t$. If two of them collide, then both are annihilated, 
since all walkers are vicious persons.
It should be noted that this {\it noncolliding condition}
seems to be very local and incidental, 
since any walker can enjoy free walking, 
while the relative distances from the
nearest-neighbor walkers are greater than two 
units of lattice spacing.
Fisher \cite{Fis84} and Huse and Fisher \cite{HF84}
derived the asymptotic form 
${\cal N}_{N}(t, \x) \sim t^{-\psi_{N}}$ in large $t$
for finite $\x$ and determined the exponent as
\begin{equation}
  \psi_{N}=\frac{1}{2} {N \choose 2} = \frac{1}{4}N(N-1).
\label{eqn:psi0}
\end{equation}
Interesting and important fact is that $\psi_{N}$ is 
nonlinear in $N$ expressing the long-ranged effect
among vicious walkers, which is a result of accumulation of
contact repulsive interactions between nearest-neighbor
walkers during the time interval $[0,t]$. Moreover, (\ref{eqn:psi0})
implies that the system possesses symmetry with respect
to permutations of the walker positions.
This hidden symmetry was clarified as explained follows.
Huse and Fisher \cite{Fis84,HF84}
mapped the enumeration problem of walks
of the $N$ particles from a set of positions
$x_{1}< \cdots < x_{N}$ to $y_{1}< \cdots < y_{N}$
in time period $t$ onto the diffusion problem of
a single particle in the $N$ dimensional space
with a set of wall restrictions (the phase space) from a position
$\x=(x_{1}, \cdots, x_{N})$ to $\y=(y_{1}, \cdots, y_{N})$
in time $t$. Assume that $p(t, y|x)$ denotes the
transition probability density of a one-dimensional
Brownian motion from $x$ to $y$ in time $t$, that is,
$p(t, y|x)=e^{-(y-x)^2/2t}/\sqrt{2 \pi t}$, then by
exploiting the method of images, they derived the
$N$-body Green function of vicious walkers in the
determinantal form \cite{det}
\begin{equation}
f(t, \y|\x) = \det_{1 \leq j, k \leq N}
[ p(t, y_{k}|x_{j})].
\label{eqn:det0}
\end{equation}
As discussed in \cite{Fis84,HF84} and explicitly shown
in \cite{KT02a}, (\ref{eqn:det0}) is found to be
factorized into a product of symmetric part
(the Schur function multiplied by Gaussian kernels)
and the antisymmetric part (the product of differences
of variables), and ${\cal N}_{N}(t,\x)$ is obtained as an 
integration of (\ref{eqn:det0}) over $\y$ with restriction
$y_{1} < y_{2} < \cdots < y_{N}$.

Based on the above knowledge on the function
${\cal N}_{N}(t,\x)$, 
let us next consider the evolution of
vicious walkers in time $t$, conditioned that they 
remain their ordering 
({\it i.e.} noncolliding condition) 
up to a given finite time $T$.
Katori and Tanemura \cite{KT02a} showed that this 
stochastic process was inhomogeneous both in space
and time and a transition in the particle
distribution was observed as time $t$ goes on from 0 to $T$.
This transition is characterized by a symmetry change,
which can be described not in the real one-dimensional
space nor the $N$-dimensional phase space
but in the space of $N \times N$ hermitian matrices. 
That is, the problem was exactly mapped to 
the statistics of $N$ real eigenvalues of $N \times N$
hermitian random matrices in a time-dependent Gaussian
ensemble. Due to the hermitian condition on
$N \times N$ matrices with complex elements
$H_{jk}=H_{jk}^{\rR}+i H_{jk}^{\rI},
i=\sqrt{-1}, 1 \leq j, k \leq N$,
$N(N+1)/2$ variables in a set 
${\cal R}=\{H_{jk}^{\rR}: 1 \leq j \leq k \leq N\}$
and $N(N-1)/2$ variables in a set
${\cal I}=\{H_{jk}^{\rI}: 1 \leq j < k \leq N\}$
are chosen as independent variables.
These $N^2$ variables in total are assumed to be 
independently distributed following the Gaussian 
distributions with zero means.
The variances of the variables in ${\cal R}$
and ${\cal I}$ are proportional to $\sigma_{\rR}$ and
$\sigma_{\rI}$, respectively, both of which
are functions of $t$.
As the time $t$ is approaching the final time $T$,
the variance $\sigma_{\rI}$ 
decreases to zero and a transition from the ensemble of
complex hermitian matrices (the Gaussian unitary ensemble, GUE)
to that of real symmetric matrices
(the Gaussian orthogonal ensemble, GOE) occurs.
By integrating over the $N'=N^2-N$ other variables
than the $N$ eigenvalues, a transition of the
eigenvalue-statistics from GUE class to
GOE class is formulated \cite{PM83}
and it is indeed realized as the time evolution of
positions of vicious walkers 
\cite{KT02a,NKT03,KKom03}.

The above results suggest the possibility 
that vicious-walker-type problems 
with $N$ walkers are generally mapped to some solvable problems
in the spaces with appropriately higher dimensions $N+N'$, in which
only the symmetries of the spaces should be considered
and the interactions (restrictions) among the original walkers
are resulted from integrating over the auxiliary
$N'$ variables.
The symmetries of the higher dimensional spaces
govern the macroscopic behaviors of the systems.
The interacting particle systems far from equilibrium
will be exactly solved, if we are able to find relevant
symmetries, which are generally hidden in the original
descriptions of the systems.

\begin{figure}
\includegraphics[width=.9\linewidth]{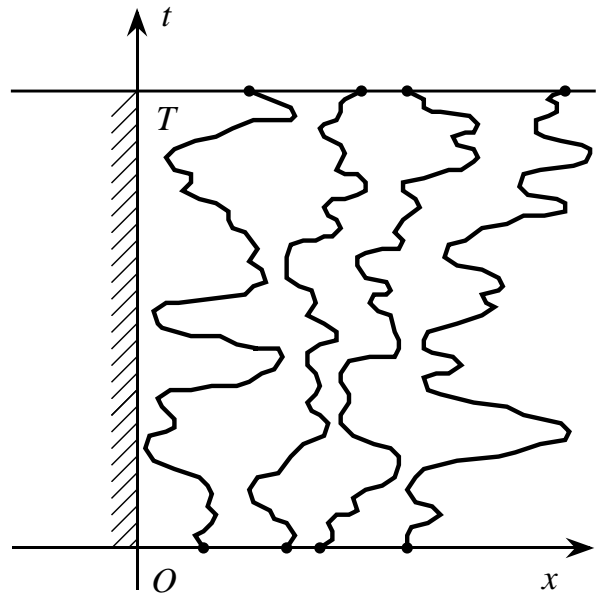}
\caption{Vicious walkers with a wall.\label{fig:fig1}}
\end{figure}

Now we propose two kinds of
problems of vicious walks, which will be
solved in the present paper in order to demonstrate
the above mentioned scheme for nonequilibrium systems.
We assume that all walkers are located
in the positive region of position as
$0 < x_{1} < x_{2} < \cdots < x_{N}$
and put an absorbing wall at the origin $\0$ (see Fig.1).
The problems are (i) to determine the probability
$\widehat{\cal N}_{N}(t,\x)$ that all walkers remain
the ordering of their positions (noncolliding condition)
with keeping apart from the wall up to time $t$, 
and (ii) to find out the 
time-dependent matrix model, whose eigenvalue-statistics 
realizes this stochastic process of vicious walkers
with a wall, and clarify the hidden symmetry transition
in the time evolution.

\begin{figure}
\includegraphics[width=.9\linewidth]{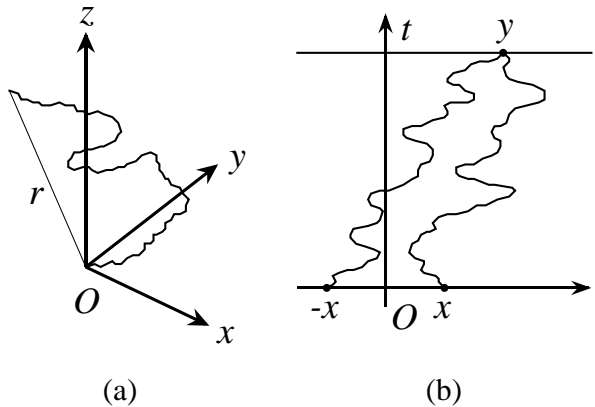}
\caption{(a) Three-dimensional Bessel process.
(b) Method of images. \label{fig:fig2}}
\end{figure}

The former problem (i) was already solved by
Krattenthaler {\it et. al} \cite{KGV00}
and the exponent governing the asymptotic form in large $t$,
$\hN(t,\x) \sim t^{-\widehat{\psi}_{N}}$,
was determined as
\begin{equation}
\widehat{\psi}_{N}=\frac{N^2}{2}.
\label{eqn:psi1}
\end{equation}
So in this paper, we will start from their result and
take the continuum limit of the model to solve the
latter problem (ii). In Section II, we construct a system of
noncolliding Brownian motions with a wall as a
diffusion scaling limit of the corresponding vicious walker model.
An important result is that
the $N$-body Green function of the obtained system
is also in the determinantal form (\ref{eqn:det0})
for $0 \leq x_{1} < \cdots < x_{N}, 
0 \leq y_{1} < \cdots < y_{N}$,
if we replace $p(t,y|x)$ by
\begin{equation}
\widehat{p}(t,y|x)=
\frac{1}{\sqrt{2 \pi t}}
\left\{ e^{-(y-x)^2/2t}-e^{-(y+x)^2/2t} \right\}.
\label{eqn:hatp}
\end{equation}
This function becomes zero as $y \to 0$, representing
the effect of the absorbing wall at $\0$.
There are two distinct ways to derive (\ref{eqn:hatp}).
(a) Consider a Brownian motion started from the origin
$\0$ in a space with dimensions $d \geq 2$.
We adopt the $d$-dimensional spherical coordinate 
${\bf r}=(r, \theta_{1}, \cdots, \theta_{d-1})$
to represent the motion. In particular, we can trace
the radial coordinate 
(the modulus of the Brownian motion) $r=r(t)$
as shown in Fig.2 (a) for $d=3$.
Since the transition probability density of $r$ is
generally described using the Bessel function,
such a stochastic process of $r$ is called the
Bessel process \cite{RY98,Yor92,BS96}.
If we multiply the transition probability density of the 
three-dimensional Bessel process by $x/y$, then
(\ref{eqn:hatp}) is obtained.
(b) For a real path of the Brownian motion from $(x,0)$ to $(y,t)$
in a spatio-temporal plane, we consider an imaginary
path from $(-x,0)$ to $(y,0)$, where $x, y > 0$
(see Fig.2 (b)).
As an analogy of electrostatic problem,
we subtract the transition probability density of the
imaginary paths from that of the real paths to obtain 
(\ref{eqn:hatp}) (the method of images).

Yor studied a temporally inhomogeneous process 
called the {\it Brownian meander}, which is obtained
as a transform of the three-dimensional Bessel process
used in the above derivation (a) of (\ref{eqn:hatp}).
He also introduced the {\it $d$-dimensional generalized meanders}
as the transform of the $d$-dimensional Bessel processes 
\cite{Yor92}.
In Section III, we will give a general theory of 
noncolliding $N$ walkers constructed as a conditioned 
system of $d$-dimensional generalized meanders. 
As a special case, it provides a proof of that
the noncolliding system of Brownian particles in the
presence of a wall is equivalent with the noncolliding
system of the Brownian meanders.
This is the complete generalization to many particle systems
with arbitrary number of particles $N$ of the fact that
the single-particle Green function (\ref{eqn:hatp}) of 
a Brownian motion with wall restriction at the origin
is proportional to the transition probability density
of a single three-dimensional Bessel process.
A key point of our proof for this new result is the proper
transform from the $N$-body Green function (\ref{eqn:det0})
to the transition probability density
by multiplying an appropriate ratio of
$\hN$'s (see Eqs. (\ref{eqn:g}) and (\ref{eqn:gNnk}) below).
Moreover, our general argument gives that, if we consider
the problem (ii) for the $d$-dimensional Bessel processes
and generalized meanders in the case of {\it even} $d$, 
it is solved using the Gaussian matrix theory with 
{\it chiral} symmetries,
which is relevant for the physics of QCD
at low energies \cite{VZ93,Ver94,JSV96,SV98}. 
Since Brownian meander is made from the {\it three-dimensional} 
Bessel process as mentioned above, it is concluded that
the present problem with (\ref{eqn:hatp})
is {\it not} related with chiral symmetry
of matrices and that matrix models in the different symmetry
classes should be considered.

The latter derivation (b) of (\ref{eqn:hatp}) gave us a
hint to find out the true symmetries, which govern the 
distribution of vicious walkers with a wall;
{\it particle-hole symmetry}.
The particle-hole symmetry is important in the BCS
theory of superconductivity.
In particular, its microscopic mean-field treatment
ignores any local interactions among particles and holes
but consider this symmetry with the so-called 
Bogoliubov-deGennes (BdG for short) Hamiltonian.
Random matrix theory of the BdG-type Hamiltonians
was introduced and developed by Altland and Zirnbauer
in order to describe the energy-level
statistics and transport properties in a metallic
quantum dot in contact with a superconductor
in a magnetic field \cite{AZ96,AZ97}.
They studied the Gaussian ensembles of the BdG Hamiltonian
in the form 
$$
{\cal H}=U^{\dagger} \left( 
\matrix{ \omega & 0 \cr 0 & -\omega} \right)
U, \quad
\omega={\rm diag}(\omega_{1}, \cdots, \omega_{N}),
$$
where $U$ is an appropriate unitary matrix.
As we will give a brief summary of their results
in Section IV, Altland and Zirnbauer discovered
four new classes of eigenvalue-statistics
in addition to the previously known three
classic Wigner-Dyson classes
(two of them are GUE and GOE, whose relation with
the vicious walks without wall was reported in
\cite{KT02a}) and three chiral symmetry classes
(two of them will be argued in Section III 
associated with the noncolliding systems of
generalized meanders constructed from
the even-dimensional Bessel processes).
In the four nonstandard symmetry classes, 
denoted by $C$, $C$I, $D$ and $D$III in 
Cartan's notation \cite{AZ97}, the classes
$C$ and $C$I are relevant in the present vicious 
walk problem.
They have the probability density functions of 
nonnegative eigenvalues in the form
\begin{eqnarray}
&& p_{\alpha, \beta}^{\rm BdG}(\omega; \sigma^2)
\propto e^{-|\omega|^2/2\sigma^2}
\prod_{1 \leq j < k \leq N}
|\omega_{k}-\omega_{j}|^{\beta}
\prod_{\ell=1}^{N} |\omega_{\ell}|^{\alpha},
\nonumber\\
\label{eqn:BdG1}
\end{eqnarray}
where $|\omega|^2=\sum_{j=1}^{N} \omega_{j}^2$ and
the indices $\alpha$ and $\beta$ are specified as
\begin{eqnarray}
&& \alpha=2, \quad \beta=2, \quad 
\mbox{for class $C$}, \nonumber\\
&& \alpha=1, \quad \beta=1, \quad 
\mbox{for class $C$I}
\nonumber
\end{eqnarray}
respectively.
We will show in Section IV that the transition
of distribution of vicious walker positions with a wall
is described by the symmetry change from the
class $C$ to the class $C$I of the BdG Hamiltonians.
This fact was already reported by Nagao \cite{Nag03},
but in this paper complementary new results will be given.
In an earlier paper \cite{KKom03}, the transition
from GUE to GOE realized in the vicious walks
without wall was characterized by the graphical
expansions with time-dependent coefficients
for the moments of walkers.
In Section V, we will introduce the M\"obius graph
expansions for the moments of the vicious walkers
with a wall. Moreover, using an exact results of
dynamical correlations by Nagao \cite{Nag03},
closed formulae for the moments will be given.
Such graphical expansions will be first reported in detail
in the present paper
for the nonstandard symmetry classes $C$ and $C$I
of Altland and Zirnbauer.
Concluding remarks are given in Section VI.

\section{VICIOUS WALK WITH A WALL AND 
ITS DIFFUSION SCALING LIMIT}

\subsection{Determinantal formula}

First we consider the $N$ independent simple and symmetric random
walks on an integer lattice 
$\Z=\{\cdots, -2, -1, 0, 1, 2, \cdots\}$ starting
from the sites $\{x_{j}\}, j=1,2, \cdots, N$, 
and denote the position of
$j$-th walker at time $n=0,1,2, \cdots$ by $x_{j}(n)$.
Assume that the initial positions are all distinct nonnegative
even integers and ordered as
$0 \leq x_{1} < x_{2} < \cdots < x_{N}$.
Then we impose the noncolliding condition up to a given 
time $m \geq 0$,
$$
  x_{1}(n) < x_{2}(n) < \cdots < x_{N}(n), \quad
n=1,2, \cdots, m.
$$
Such conditional walks are called {\it vicious walks up
to time} $m$ \cite{Fis84}. Here we impose further
restriction on the walks as
$$
  x_{j}(n) \geq 0, \quad j=1,2, \cdots, N, \quad
n=1,2, \cdots, m.
$$
In other words, there is a wall at the origin and
all the walkers are conditioned {\it never to collide with
each other nor to collide with the wall during the
time interval} $0 \leq n \leq m$. 

Let $\widehat{N}_{N}(m, \y|\x)$,
$\x=(x_{1}, \cdots, x_{N}), \y=(y_{1}, \cdots, y_{N})$,
be the total number
of the vicious walks with wall restriction, in which the
$N$ walkers start from the positions $x_{j}, j=1,2, \cdots, N$,
and arrive at the positions $y_{j}, j=1,2, \cdots, N$,
at time $m$. 
Krattenthaler {\it et al.} gave the determinantal formula
to this number as \cite{KGV00}
\begin{eqnarray}
&& \widehat{N}_{N}(m, \y|\x) \nonumber\\
&=& \det_{1 \leq j, k \leq N} 
\left[
{ m \choose \frac{m+x_{j}-y_{k}}{2} } 
- { m \choose \frac{m+x_{j}+y_{k}+2}{2} }
\right].
\nonumber
\end{eqnarray}
Suppose that all random walks start from given initial 
positions $\x$. Since the total number of walks is 
$2^{mN}$, the probability that they are vicious walks
with wall restriction and end up with 
positions $\y$ is $\widehat{N}_{N}(m, \y|\x)/2^{mN}$.

\subsection{Diffusion scaling limit}

In order to take the continuum limit of the vicious walks
to derive the system of noncolliding Brownian motions,
we introduce a function
$
\phi_{L}(x)=2 [Lx/2],
$
for $L > 0, x \in \R$ (the set of all real numbers),
where $[z]$ denotes the largest integer not greater than $z$,
and let $\phi_{L}(\x)=(\phi_{L}(x_{1}), \cdots, \phi_{L}(x_{N}))$.
By Stirling formula, we can take the diffusion scaling
limit as
$$
\lim_{L \to \infty} \left(\frac{L}{2}\right)^N
\frac{1}{2^{mN}}
\widehat{N}_{N}(\phi_{L^2}(t), \phi_{L}(\y)|\phi_{L}(\x))
= \f(t,\y|\x)
$$
for $\x, \y$ with
$0 \leq x_{1}< \cdots < x_{N}, 0 \leq y_{1} < \cdots < y_{N}$,
where
\begin{equation}
\f(t, \y|\x) = \det_{1 \leq j, k \leq N}
\left[ \widehat{p}(t, y_{k}|x_{j}) \right]
\label{eqn:fN}
\end{equation}
with (\ref{eqn:hatp}).
Let $T>0$ and consider a system of $N$ Brownian motions
conditioned never to collide with each other nor to collide 
with the wall at $x=0$ in $[0,T]$. Set
\begin{equation}
\hN(t, \x)= 
\int_{0 \leq y_{1} < \cdots < y_{N}}
d \y \f(t, \y|\x),
\label{eqn:calN}
\end{equation}
where $d \y=\prod_{j=1}^{N} dy_{j}$.
Then the transition probability density
from the state $0 \leq x_{1} < \cdots < x_{N}$
at time $s$ to the state $0 \leq y_{1} < \cdots < y_{N}$
at time $t (\geq s)$ of such a system is given by 
\begin{equation}
\gT(s,\x; t, \y)=
\frac{\f(t-s, \y|\x) 
\hN(T-t, \y)}
{\hN(T-s, \x)},
\label{eqn:g}
\end{equation}
since the numerator is the probability
that we have the noncolliding (with each other and with a wall)
Brownian paths from $\x$ at time $s$ to $\y$ at time $t$
{\it and} these paths keep noncolliding from the
time $t$ up to time $T$ as well, and
the denominator is the probability
that the Brownian paths are noncolliding all
during time interval $[s, T]$.

It is useful to rewrite (\ref{eqn:fN}) as
\begin{eqnarray}
&& \f (t, \y |\x ) = (2 \pi t)^{-N/2} 
{\rm sp}_{\xi(\y)}\left(e^{x_{1}/t}, \cdots, e^{x_{N}/t}\right)
\nonumber\\
&\times& e^{-(|\x|^2+|\y|^2)/2t}
\prod_{j=1}^{N}(e^{x_{j}/t}-e^{-x_{j}/t}) \nonumber\\
&\times&
\prod_{1 \leq i < j \leq N}
\Big\{ (e^{x_{j}/t}-e^{x_{i}/t})(e^{(x_{i}+x_{j})/t}-1)\Big\}
\nonumber\\
&\times& \Big\{\prod_{j=1}^{N} e^{x_{j}/t}
\Big\}^{-N+1},
\label{eqn:f2}
\end{eqnarray}
where $\xi(\y)=(\xi_{1}(\y), \cdots, \xi_{N}(\y))$
with $\xi_{j}(\y)=y_{N-j+1}-(N-j+1), j=1,2, \cdots, N$,
$$
{\rm sp}_{\xi}(z_{1}, \cdots, z_{N})=
\frac{\det(z_{i}^{\xi_{j}+N-j+1}-z_{j}^{-(\xi_{j}+N-j+1)})}
{\det(z_{i}^{N-j+1}-z_{i}^{-(N-j+1)})},
$$
and $|\x|^2=\sum_{j=1}^{N} x_{j}^2$.
Remark that ${\rm sp}_{\lambda}(z_{1}, \cdots, z_{N})$ is 
the character of the irreducible representation 
corresponding to a partition $\lambda$
of the symplectic Lie algebra 
(see, for example, Lectures 6 and 24 in \cite{FH91}). 
Since we know the formula
$$
{\rm sp}_{\xi}(1, \cdots, 1)=
\prod_{1 \leq i < j \leq N} 
\frac{\ell_{j}^2-\ell_{i}^{2}}{m_{j}^{2}-m_{i}^{2} }
\prod_{j=1}^{N} 
\frac{\ell_{j}}{m_{j}}
$$
with
$\ell_{j}=\xi_{j}+N-j+1$, $m_{j}=N-j+1$
\cite{FH91}
and the integral
((17.6.6) on page 354 in \cite{Meh91})
\begin{eqnarray}
&& \int d \x e^{- |\x|^{2}/2}
\prod_{1 \leq i < j \leq N}
|x_{j}^{2}-x_{i}^{2}|^{2 \gamma}
\prod_{j=1}^{N} |x_{j}|^{2 a-1} \nonumber\\
&=& 2^{aN + \gamma N (N-1)}
\prod_{j=1}^{N} \frac{\Gamma(1+j \gamma) 
\Gamma(a+\gamma(j-1))}{\Gamma(1+\gamma)},
\nonumber
\end{eqnarray}
we have the asymptotic 
\begin{equation}
\hN(t, \x)=\frac{1}{\widetilde{c}_{N}}
\h(\x/\sqrt{t})\left(
1+{\cal O}(\x/\sqrt{t})\right)
\label{eqn:asym}
\end{equation}
for $|\x|/\sqrt{t} \to 0$, where
$\widetilde{c}_{N}=(\pi/2)^{N/2} \prod_{j=1}^{N}\Gamma(2j)/
\Gamma(j)$ with the Gamma function $\Gamma(z)$ and
\begin{equation}
\h(\x)=\prod_{1 \leq j < k \leq N}(x_{k}^{2}-x_{j}^{2}) 
\prod_{\ell=1}^{N} x_{\ell}.
\label{eqn:hath}
\end{equation}
Substituting eqs.(\ref{eqn:f2}) and (\ref{eqn:asym}) 
into eq.(\ref{eqn:g}), we find
\begin{eqnarray}
&& \gT (0, \0; t, \y) = \widehat{c}_{N}T^{N^2/2}t^{-N(2N+1)/2} 
\nonumber\\
&& \qquad \times
e^{-|\y|^2/2t} \h(\y) \hN (T-t,\y),
\label{eqn:g0}
\end{eqnarray}
where $\widehat{c}_{N}=1/\prod_{j=1}^{N} \Gamma(j)$.
The transition probability densities 
(\ref{eqn:g}) and (\ref{eqn:g0}) define
the {\it $N$ noncolliding Brownian motions with 
wall restriction} in time interval $(0,T]$ \cite{KT02b}.

It should be noted that $\hN(t, \x)$ is the
noncolliding probability of Brownian motions
with wall restriction
and (\ref{eqn:asym}) gives the power-law behavior
$\hN(t,\x) \sim t^{-\hat{\psi}_{N}}$
in large $t$ for finite $\x$ with the critical exponent
(\ref{eqn:psi1}).

\subsection{Transition from class $C$ to class $C$I}

By (\ref{eqn:asym}), the $T \to \infty$ limit of
(\ref{eqn:g}) is determined and simply given as
\begin{eqnarray}
&& \p(0, \x; t, \y) \equiv
\lim_{T \to \infty} \gT(0, \x; t, \y) \nonumber\\
&& \qquad = \frac{\h(\y)}{\h(\x)} \f(t, \y|\x),
\label{eqn:phat}
\end{eqnarray}
where we have set $s=0$ and used (\ref{eqn:hath}).
Moreover, we can take the $\x \to \0$ limit of
(\ref{eqn:phat}) to obtain
\begin{equation}
\p (0, \0; t, \y)
= \widehat{c^{\prime}}_N
t^{-N(2N+1)/2} e^{-|\y|^2/2t}\h(\y)^2,
\label{eqn:phat0}
\end{equation}
where 
$\widehat{c^{\prime}}_{N}=(2/\pi)^{N/2}
/\prod_{j=1}^N \Gamma(2j)$.
That is, we have the identity
$$
\p(0,\0; t, \y)=N! p^{\rm BdG}_{2,2}(\y; t)
$$
for $0 \leq y_{1} < \cdots < y_{N}$, where
$p^{\rm BdG}_{2,2}(\omega; \sigma^2)$ is the probability
density function (\ref{eqn:BdG1}) of 
nonnegative eigenvalues of
the BdG Hamiltonian in the class $C$ ($\alpha=\beta=2$).
On the other hand, if we set $t=T$ in (\ref{eqn:g}),
for $\hN(0, \y)=1$, we have
\begin{equation}
\gT(0, \0; T, \y) = \widehat{c}_{N}
t^{-N(N+1)/2} e^{-|\y|^2/2t} \h(\y),
\label{eqn:eqadd1}
\end{equation}
which implies the identity
$$
\gT(0, \0; T, \y)=N! 
p^{\rm BdG}_{1,1}(\y; t).
$$
That is, at $t=T$, $\gT(0, \0; T, \y)$
is identified with the 
probability density function (\ref{eqn:BdG1}) of 
nonnegative eigenvalues of
the BdG Hamiltonian in the class $C$I ($\alpha=\beta=1$).

The above results mean the following facts.
If we consider the $N$ noncolliding Brownian motions
with wall restriction up to a finite time $T>0$, 
in which all particles start from the origin,
as the ratio $t/T \to 0$, the distribution of
particle positions is asymptotically described by the
eigenvalue-statistics of the BdG Hamiltonian
in the class $C$. On the other hand, at the final time
$t=T$, it can be identified with the eigenvalue-statistics
of the BdG Hamiltonian in the class $C$I.
There occurs, thus, a transition 
from the class $C$ distribution to the class $C$I
distribution as time $t$ goes on from 0 to $T$
in our stochastic process.

The essential difference between (\ref{eqn:phat0}) 
and (\ref{eqn:eqadd1}) is found in the exponents
of the factors $\h(\y)^{\beta}$ such that
$\beta=2$ in (\ref{eqn:phat0}) and $\beta=1$
in (\ref{eqn:eqadd1}). This factor expresses 
strong repulsive interactions among particles
and between the wall and each particle, in which
the larger exponent $\beta$ gives stronger repulsion
for short distances.
At the very early stage of the process, $t/T \ll 1$,
the repulsion may be strong, since the noncolliding condition
will be imposed for a long time period up to time $T$
in the future. As the time $t$ goes on, the repulsion
strength is decreasing as is the remaining time until $T$,
and attains its minimum at $t=T$.

\subsection{Stochastic differential equations}

As explained in \cite{KT02a} in the case without wall restriction,
the positions $\x(t)=(x_{1}(t), \cdots, x_{N}(t))$
of the $N$ noncolliding Brownian motions solve the stochastic
differential equations in a modified type of
Dyson's Brownian motion model \cite{Dys62,Meh91}.
In the present case with wall restriction, we have
$$
dx_{j}(t)=\widehat{E}_{j}^{T}(\x(t)) dt + dB_{j}(t),
$$
for $0 \leq x_{1} < \cdots < x_{N}, 0 < t \leq T$,
where 
$$
\widehat{E}_{j}^{T}(\x)=\frac{\partial}{\partial x_{j}}
\ln \hN(T-t; \x), 
$$
and $\{B_{j}(t)\}_{j=1}^{N}$ are $N$ independent standard Brownian
motions
\begin{eqnarray}
  && B_{j}(0)=0, \quad
\langle B_{j}(t) \rangle =0, \nonumber\\
&& \left\langle \left(
B_{j}(t)-B_{j}(s) \right) 
\left(B_{k}(t)-B_{k}(s) \right) \right\rangle
=|t-s|\delta_{jk},
\nonumber
\end{eqnarray}
for any $t, s > 0, j,k=1,2, \cdots, N$.
In particular, in the limit $T \to \infty$, 
(\ref{eqn:asym}) with (\ref{eqn:hath}) gives
the equations
\begin{eqnarray}
&& dx_{j}(t) = dB_{j}(t) + \frac{1}{x_{j}(t)} dt \nonumber\\
&+& \sum_{1 \leq k \leq N, k \not=j}
\left\{ \frac{1}{x_{j}(t)-x_{k}(t)}+
\frac{1}{x_{j}(t)+x_{k}(t)}
\right\} dt \qquad
\label{eqn:eqadd2}
\end{eqnarray}
for $1 \leq j \leq N$ \cite{KT02b}.
In the stochastic differential equation for the position
of $j$-th particle, the drift terms
$dt/x_{j}(t)$ and $dt/(x_{j}(t)-x_{k}(t))$ represent
repulsive force from the wall at the origin and
that from the $k$-th particle, respectively.
In addition to them, there are the terms in the form
$dt/(x_{j}(t)+x_{k}(t)), 1 \leq k \leq N, k \not=j$,
which can be interpreted as effective repulsive
forces from the mirror images of other particles
located at $-x_{k}(t), 1 \leq k \leq N, k \not=j$.

\section{NONCOLLIDING MEANDERS
AND CHIRAL RANDOM MATRIX THEORY}

\subsection{Definitions of elementary processes}

Consider a diffusion equation in dimension $d \geq 2$
$$
\frac{\partial}{\partial t} u(t, \y|\x)
=\frac{1}{2} \sum_{j=1}^{d} \frac{\partial^2}{\partial y_{j}^2}
u(t, \y|\x)
$$
with the initial condition 
$u(0, \y|\x)=\delta(\x-\y)$.
We use the spherical coordinates $\x=(x, \theta_{1}, \cdots, \theta_{d-1}),
\y=(y, \varphi_{1}, \cdots, \varphi_{d-1})$ and integrate
over all the angular variables to obtain a differential
equation for the radial coordinate (the modulus)
$$
\frac{\partial}{\partial t} \bar{u}(t, y|x)
= \frac{1}{2} \left[ \frac{\partial^2}{\partial y^2}
+ \frac{d-1}{y} \frac{\partial}{\partial y} \right]
\bar{u}(t, y|x).
$$
The unique solution of this equation satisfying the
initial condition $\bar{u}(0,y|x) y^{d-1} dy=
\delta(x-y) dy$ for $x>0$ is given as
$$
\bar{u}(t,y|x)=\frac{1}{(xy)^{\nu}} \frac{1}{t}
e^{-(x^2+y^2)/2t} I_{\nu}\left(\frac{xy}{t}\right),
$$
where $\nu=(d-2)/2$ and 
$I_{\nu}(z)$ is the modified Bessel function
$$
I_{\nu}(z)= \sum_{n=0}^{\infty} \frac{
(z/2)^{2n+\nu}}{ n ! \Gamma(\nu+n+1)}.
$$
If we set $p^{(\nu)}(t,y|x)=\bar{u}(t,y|x) y^{d-1}$,
then it is normalized as
$\int_{0}^{\infty} p^{(\nu)}(t,y|x) dy=1$
for any $x>0$. We define $p^{(\nu)}(t,y|0)$ by
the $x \to 0$ limit of $p^{(\nu)}(t,y|x)$.
Then we have
\begin{eqnarray}
p^{(\nu)}(t, y|x)
&=&
\frac{y^{\nu+1}}{x^{\nu}}
\frac{1}{t} e^{-(x^2+y^2)/2t}
I_{\nu} \left( \frac{x y}{t} \right),
\, x>0, y \ge 0,
\nonumber
\\
p^{(\nu)}(t, y|0)
&=&
\frac{y^{2\nu +1}}
{2^{\nu} \Gamma (\nu +1)t^{\nu+1}} e^{-y^2 /2t},
\quad y \ge 0.
\label{eqn:Bessel}
\end{eqnarray}
The $d=2(\nu+1)$ dimensional Bessel process
is defined so that its transition probability density is 
given by (\ref{eqn:Bessel}) \cite{RY98,Yor92,BS96}.

For $0 \leq u \leq T, w \geq 0$, $\kappa \in [0, 2(\nu+1) )$, 
we consider 
\begin{equation}
h^{(\nu,\kappa)}_{T}(u,w)
= \int_0^\infty dz \ p^{(\nu)}(T-u, z|w) z^{-\kappa}.
\label{eqn:h}
\end{equation}
That is, we multiply a weight $z^{-\kappa}$ at
the final time $T$, so that, as the arrival position
$z$ is nearer to the origin, the path is more enhanced.
Then the transition probability density from $x$ at time $s$
to $y$ at time $t$ with such bias at time $T$,
$ 0 \le s< t \le T$, is given by
\begin{equation}
p^{(\nu,\kappa)}_{T}(s,x; t,y)
= \frac{1}{h^{(\nu,\kappa)}_{T}(s,x)}
p^{(\nu)}(t-s, y|x) h^{(\nu,\kappa)}_{T}(t,y)
\label{eqn:meander}
\end{equation}
for $ x, y \geq 0 $.
This bias makes the process defined by
(\ref{eqn:meander}) be temporally inhomogeneous,
and Yor called it the {\it generalized meander}
indexed $(\nu, \kappa)$.
In particular, when $\nu=1/2$ and $\kappa =1$,
the process is called the {\it Brownian meander}
\cite{Yor92}.
The transformation from (\ref{eqn:Bessel}) to 
$p^{(\nu,\kappa)}_{T}(s,x; t,y)$ by (\ref{eqn:meander})
is a generalization of the $h$-transform of
Doob \cite{Do84}.
Remark that, if we replace $T$ by $u$ in (\ref{eqn:h}),
then $h_{u}^{(\nu, \kappa)}(u,w)=w^{-\kappa}$,
since $p^{(\nu)}(0, z|w)=\delta(z-w)$.
In the case $\nu=1/2, \kappa=1$ and $s=0$,
(\ref{eqn:meander}) becomes, by this replacement,
$p^{(1/2)}(t, y|x) \times x/y$, which is equal to
(\ref{eqn:hatp}) 
for $I_{1/2}(z)=\sqrt{2/(\pi z)} \sinh z$.
This is the derivation (a)
of (\ref{eqn:hatp}) mentioned in Section I.

\subsection{Non-colliding systems and rectangular
random matrices}

Now we consider a system of $N$ generalized meanders
conditioned that they never collide with each other 
for a time interval $(0, T], T> 0$. 
Using the determinantal formula
in (\ref{eqn:det0}) and following the same way
as (\ref{eqn:g}) and (\ref{eqn:meander}),
the transition probability density is given by
\begin{equation}
g_{N,T}^{(\nu, \kappa)}(s, {\bf x}; t, {\bf y})
= \frac{f_{N,T}^{(\nu,\kappa)}(s, {\bf x}; t,{\bf y})
{\cal N}_{N,T}^{(\nu, \kappa)}(T-t, {\bf y})}
{{\cal N}_{N,T}^{(\nu, \kappa)}(T-s, {\bf x})}
\label{eqn:gNnk}
\end{equation}
for $0\le s< t \le T,
0 \leq x_{1} < \cdots < x_{N},
0 \leq y_{1} < \cdots < y_{N}$,
where
\begin{equation}
f_{N,T}^{(\nu,\kappa)}(s, {\bf x}; t, {\bf y}) =
\det_{1 \leq j, k \leq N} \left[
p^{(\nu,\kappa)}_{T}(s,x_j; t,y_k)
\right],
\label{eqn:fNnu}
\end{equation}
and
$$
{\cal N}_{N,T}^{(\nu, \kappa)}(T-t, {\bf x})
= \int_{0 \leq y_{1} < \cdots < y_{N}} d {\bf y}
f_{N,T}^{(\nu,\kappa)}(T-t,{\bf x}, T, {\bf y}).
$$

Since $f_{N,T}^{(\nu,0)}(s, \x; t, \y)$ is temporally homogeneous
and independent of $T$, we will write it as
$f_{N}^{(\nu)}(t-s , {\bf y} |{\bf x})$.
Moreover, note that 
\begin{equation}
f_{N,T}^{(\nu,\kappa)}(s, {\bf x}; t, {\bf y}) =
\frac{1}{h^{(\nu,\kappa)}_{T}(s,{\bf x})}
f_{N}^{(\nu)}(t-s, {\bf y}|{\bf x}) 
h^{(\nu,\kappa)}_{T}(t,{\bf y}),
\label{eqn:trans}
\end{equation}
where
$h^{(\nu,\kappa)}_{T}(t,{\bf x}) 
= \prod_{j=1}^N h^{(\nu,\kappa)}_{T}(t,x_j)$,
and that
$h^{(\nu, \kappa)}_{T}(T, {\bf x})
= \prod_{j=1}^{N} x_{j}^{-\kappa}$.
Then (\ref{eqn:gNnk}) can be written as
\begin{eqnarray}
&&g_{N,T}^{(\nu, \kappa)}(s, {\bf x}; t, {\bf y}) \nonumber\\
&=& \frac{1}{\widetilde{{\cal N}}_{N}^{(\nu, \kappa)}(T-s, {\bf x})}
f_{N}^{(\nu)}(t-s, {\bf y}|{\bf x})
\widetilde{{\cal N}}_{N}^{(\nu, \kappa)}(T-t, {\bf y})
\qquad 
\label{eqn:gNnk1}
\end{eqnarray}
with
\begin{equation}
\widetilde{{\cal N}}_{N}^{(\nu, \kappa)}(t, {\bf x})
= \int_{0 \leq y_{1} < \cdots < y_{N}} d {\bf y} \
f_{N}^{(\nu)}(t, {\bf y}|{\bf x}) 
\prod_{j=1}^{N} y_{j}^{-\kappa}.
\label{eqn:tildeN}
\end{equation}
The important point is that
we can confirm that (\ref{eqn:gNnk1}) 
with $\nu=1/2$ and $\kappa=1$ is
equal to the transition probability density
(\ref{eqn:g}) of the noncolliding Brownian motions 
with a wall.
In other words, we found the equivalence between
{\it the noncolliding Brownian motions with a wall}
and {\it the noncolliding Brownian meanders}.
If we set $N=1$ in (\ref{eqn:tildeN}) with $\nu=1/2,
\kappa=1$, we have
$\widetilde{{\cal N}}_{1}^{(1/2,1)}(t,x)=1/x$.
Then (\ref{eqn:gNnk1}) with $s=0$ is reduced to the equality
$\widehat{p}(t,y|x)=p^{(1/2)}(t,y|x) \times x/y$,
which is the statement given as the derivation (a)
of (\ref{eqn:hatp}) in Section I.

We then consider the limit $|{\bf x}| \to 0$ to define the
noncolliding generalized meanders all started from the
origin $\0$ at the initial time $s=0$
for nonnegative integers $\nu$ ({\it i.e.} even $d$).
Consider an arbitrary $N_{1} \times N_{2}$ complex matrices
with $N_{1} \geq N_{2}$.
We denote by ${\cal M}(N_{1}, N_{2}; \C)$ the space of all
such matrices.
It is known \cite{Hua63} 
that any matrix $A \in {\cal M}(N_{1}, N_{2}; {\bf C})$
can be expressed by
\begin{equation}
  A = U^{\dagger} \Lambda V,
\label{eqn:polar1}
\end{equation}
where $U$ and $V$ are unitary matrices with sizes $N_{1}$ and
$N_{2}$, respectively, and $\Lambda$ is the $N_{1} \times N_{2}$
matrix in the form
\begin{equation}
  \Lambda = \left( \matrix{ \widehat{\Lambda} \cr \0} \right)
\quad
\mbox{with } \,
\widehat{\Lambda}={\rm diag}
(a_{1}, a_{2}, \cdots, a_{N_{2}}),
\label{eqn:polar2}
\end{equation}
and where $a_{j} \geq 0, 1 \leq j \leq N_{2}$.
The matrices $(U, V)$ parameterize the coset space
$U(N_{1}) \times U(N_{2})/[U(1)]^{N_{2}}$, where
$[U(1)]^{N_{2}}$ is the diagonal subgroup of $U(N_{2})$, and thus
the $(U, \Lambda, V)$ can be regarded as ``spherical coordinates"
in the space ${\cal M}(N_{1}, N_{2}; \C)$. It should be remarked
that $\{a_{1}, \cdots, a_{N_{2}}\}$ are not eigenvalues of $A$;
they will be referred to as ``radial coordinates".
The following integral formula proved in \cite{JSV96} is useful.
Let $d \mu(U,V)$ be the Haar measure of 
$U(N_{1}) \times U(N_{2})/[U(1)]^{N_{2}}$.
For $A, B \in {\cal M}(N_{1}, N_{2}; {\bf C})$,
set $A=U^{\dagger}_{A} \Lambda_{A} V_{A}$, 
$B=U^{\dagger}_{B} \Lambda_{B} V_{B}$, where
$U_{A}, U_{B} \in U(N_{1})$,
$V_{A}, V_{B} \in U(N_{2})/[U(1)]^{N_{2}}$,
$$
\Lambda_{A}=\left( \matrix{\widehat{\Lambda}_{A} \cr \0} \right),
\quad
\Lambda_{B}=\left( \matrix{\widehat{\Lambda}_{B} \cr \0} \right),
$$
with $\widehat{\Lambda}_{A}={\rm diag}(a_{1}, \cdots, a_{N_{2}})$,
$\widehat{\Lambda}_{B}={\rm diag}(b_{1}, \cdots, b_{N_{2}})$,
$a_{j} \geq 0, b_{j} \geq 0, 1 \leq j \leq N_{2}$.
Then for an arbitrary constant $\sigma$,
\begin{eqnarray}
&& \int d \mu(U_{A}, V_{A}) \
\exp \left( -\frac{1}{2 \sigma^2} {\rm tr} \{
(A-B)^{\dagger} (A-B) \} \right) \nonumber\\
&\propto&
 \frac{\det_{1 \leq j, k \leq N_{2}}
\left[ \exp \left(-\frac{a_{j}^2+b_{k}^2}{2 \sigma^2} \right)
I_{N_{1}-N_{2}} \left( \frac{a_{j} b_{k}}{\sigma^2} \right)
\right]}
{\displaystyle{
\prod_{j=1}^{N_{2}} (a_{j} b_{j})^{N_{1}-N_{2}}
\prod_{1 \leq j < k \leq N_{2}} (a_{j}^2-a_{k}^2)
(b_{j}^2-b_{k}^2)}}.
\nonumber
\end{eqnarray}
Remark that this integral formula can be regarded as 
a version of the Harish-Chandra (Itzykson-Zuber) formula 
\cite{HC57,IZ80,Meh81}.

Using this integral formula, (\ref{eqn:fNnu}) with 
a nonnegative integer $\nu$ and $\kappa=0$
is written as
\begin{eqnarray}
&& f_{N}^{(\nu)}(t, {\bf y}|{\bf x}) \nonumber\\
&\propto&
\prod_{1 \leq j < k \leq N} (x_{j}^2-x_{k}^2)
\prod_{j=1}^{N} y_{j}^{2\nu+1} \prod_{1 \leq j < k \leq N}
(y_{j}^2-y_{k}^2) \nonumber\\
&\times& \int d\mu (U_{Y}, V_{Y}) \
\exp \left( -\frac{1}{2t} {\rm tr} \{(X-Y)^{\dagger} (X-Y)\} \right).
\nonumber
\end{eqnarray}
Since ${\rm tr} \{(X-Y)^{\dagger} (X-Y)\} \to {\rm tr} Y^{\dagger} Y
= |{\bf y}|^2$ as $|{\bf x}| \to 0$,
we have
\begin{eqnarray}
&& \lim_{|{\bf x}| \to 0} 
\frac{f_{N}^{(\nu)}(t, {\bf y}|{\bf x})}
{\displaystyle{\prod_{1 \leq j < k \leq N} (x_{j}^2-x_{k}^2)}}
\nonumber\\
&\propto& \prod_{j=1}^{N} y_{j}^{2\nu+1}
\prod_{1 \leq j < k \leq N} (y_{j}^2-y_{k}^2) 
e^{-|{\bf y}|^2/2t}.
\label{eqn:xlim0}
\end{eqnarray}
Above argument proves the following result.
Let $\nu$ be a nonnegative integer and
$0 \leq \kappa < 2(\nu+1)$.
The limit $|{\bf x}| \to 0$ of
$g_{N,T}^{(\nu, \kappa)}(0, {\bf x}, t, {\bf y})$ is given by
\begin{eqnarray}
&& g_{N,T}^{(\nu, \kappa)}(0, \0; t, {\bf y}) 
= c \ e^{-|{\bf y}|^2/2t} \nonumber\\
&& \times  \prod_{j=1}^{N} y_{j}^{2\nu+1}
\prod_{1 \leq j < k \leq N} (y_{j}^2-y_{k}^2) \
\widetilde{{\cal N}}_{N}^{(\nu, \kappa)}
(T-t, {\bf y}), \qquad
\label{eqn:gNnk2}
\end{eqnarray}
where $c$ is a normalization 
constant determined by
$\int_{0 \leq y_{1} < \cdots < y_{N}} d\y
g_{N,T}^{(\nu, \kappa)}(0, \0; t, {\bf y}) =1$.
The $N$ noncolliding generalized meanders {\it all started
from the origin $\0$ at time 0} is defined by the
transition probability density (\ref{eqn:gNnk2}).

\subsection{Chiral Gaussian ensembles
and transition of chiral symmetries}

For the space ${\cal M}(N_{1}, N_{2}; \C)$ of all
$N_{1} \times N_{2}$ complex matrices, we introduce
the integration measure
$
d v(A) = \prod_{j=1}^{N_{1}} \prod_{k=1}^{N_{2}}
d A_{jk}^{\rR} d A_{jk}^{\rI}
$
for $A =(A_{jk}) \in {\cal M}(N_{1}, N_{2}; \C)$
with $A_{jk}=A_{jk}^{\rR}+i A_{jk}^{\rI}, i=\sqrt{-1}$.
The chiral Gaussian unitary ensemble (chGUE for short) with variance 
$\sigma^2$ is the ensemble of matrices 
$A \in {\cal M}(N_{1}, N_{2}; \C)$ with the probability
measure
$$
d \mu^{\rm chGUE}(A; \sigma^2)
\propto \exp \left( - \frac{1}{2 \sigma^2}
{\rm tr} A^{\dagger} A \right) dv (A).
$$
For $A \in {\cal M}(N_{1}, N_{2}; \C)$ with the
polar coordinates (\ref{eqn:polar1}) and (\ref{eqn:polar2}),
we can show that
the probability density function of the ``radial coordinates"
${\bf a}=(a_{1}, \cdots, a_{N_{2}})$ of 
$A \in {\cal M}(N_{1}, N_{2}; \C)$
in chGUE with variance $\sigma^2$ is given as
\begin{eqnarray}
&& p^{\rm chGUE}({\bf a}; \sigma^2) \nonumber\\
&\propto&
e^{-|\a|^2/2\sigma^2}
\prod_{j=1}^{N_{2}} a_{j}^{2(N_{1}-N_{2})+1}
\prod_{1 \leq j < k \leq N_{2}} (a_{j}^2-a_{k}^2)^2.
\nonumber
\end{eqnarray}
Next we set ${\cal M}(N_{1}, N_{2}; \R)$ be the space of
all $N_{1} \times N_{2}$ {\it real} matrices for $N_{1} \geq N_{2}$.
The chiral Gaussian orthogonal ensemble (chGOE) with variance 
$\sigma^2$ is the ensemble of matrices 
$B \in {\cal M}(N_{1}, N_{2}; \R)$ with the probability
measure
$$
d \mu^{\rm chGOE}(B; \sigma^2)
\propto \exp \left( - \frac{1}{2 \sigma^2}
{\rm tr} B^{T} B \right) dv' (B)
$$
with $dv'(B)=\prod_{j=1}^{N_{1}}\prod_{k=1}^{N_{2}} dB_{jk}$.
The probability density function of the radial coordinates 
${\bf b}=(b_{1}, \cdots, b_{N_{2}})$ is given in the form
$$
p^{\rm chGOE}({\bf b}; \sigma^2) 
\propto e^{-|\b|^2/2\sigma^2}
\prod_{j=1}^{N_{2}} b_{j}^{N_{1}-N_{2}} 
\prod_{1 \leq j < k \leq N_{2}} |b_{j}^2-b_{k}^2|.
$$
Then we consider the distribution of the sum of two rectangular
matrices
$
   C=A+B,
$
in which $A$ and $B$ are chosen from chGUE and chGOE, respectively.
The distribution function of $C$ is the convolution of
those of chGUE and chGOE.
Consider the ensemble of matrices 
$C \in {\cal M}(N_{1}, N_{2}; {\bf C})$,
in which the probability measure is given as
\begin{eqnarray}
&& d \mu^{\rm chGUE/GOE}(C; \sigma_{1}^{2}, \sigma_{2}^{2})
\nonumber\\
&=& \int_{B \in {\cal M}(N_{1}, N_{2}; {\bf R})}
d\mu^{\rm chGUE}(C-B; \sigma_{1}^{2}) 
d \mu^{\rm chGOE}(B; \sigma_{2}^{2}).
\nonumber
\end{eqnarray}
We denote the probability density function of the radial coordinates
${\bf c}=(c_{1}, \cdots, c_{N_{2}})$ of matrix
$C$ in this ensemble by
$p^{\rm chGUE/GOE}({\bf c}; \sigma_{1}^{2}, \sigma_{2}^{2}; N_{1}, N_{2})$.

Comparing the above definitions and (\ref{eqn:gNnk2}),
we can prove the following equality for
nonnegative integers $\nu$,
\begin{eqnarray}
&& g_{N,T}^{(\nu,\nu +1)}(0, \0; t, {\bf y}) \nonumber\\
&=& N ! \ p^{\rm chGUE/GOE} \left({\bf y}; 
t \left( 1- \frac{t}{T} \right), \frac{t^2}{T};
N+\nu, N \right), \nonumber
\label{eqn:chiraltransition}
\end{eqnarray}
where $0 \leq y_{1} < \cdots < y_{N}$.
It implies that, if $\nu=0,1,2, \cdots$,
the time-evolution of the
noncolliding generalized meanders indexed $(\nu, \nu+1)$
is represented by the transition of the
eigenvalue-statistics from the chGUE class to the 
chGOE class.

\section{GAUSSIAN ENSEMBLES OF BOGOLIUBOV-DEGENNES
RANDOM MATRICES}

Since we have found that the noncolliding Brownian motion
with a wall is equivalent with the noncolliding system
of Brownian meanders with indices $\nu=1/2$ and $\kappa=1$,
it does not belong to the chiral symmetry classes
discussed in the previous section.
We have to consider the ensembles of hermitian
matrices in the form of the Bogoliubov-deGennes
Hamiltonian for the mean-field theory of superconductivity.

As shown in II-C, the time evolution of spatial distribution
of particles in the present system can be regarded as
a transition of the eigenvalue-statistics of the BdG
random Hamiltonian in the class $C$ to the class $C$I.
The former statistics is characterized by the exponent
$\beta=2$ of the repulsive factor $\h(\y)^{\beta}$
and the latter by $\beta=1$. It does not imply,
however, that the functional form of the distribution is
maintained in the form (\ref{eqn:BdG1}) with $\alpha=\beta$
and only the exponent $\beta$ changes continuously as the time
passes. It this section, using a version of 
Harish-Chandra (Itzykson-Zuber) integral formula over
unitary group, we show the fact that the time evolution
of the present process is described by a {\it two-matrix
model} coupling random matrices, one of which is chosen
from a Gaussian ensemble of the BdG Hamiltonian
matrices of class $C$ and other of which is from that of
class $C$I. There the time dependence of variances of these
two ensembles is different from each other.

\subsection{Hermitian and real symmetric matrices
with particle-hole symmetry}

We consider the space of the hermitian matrices specified
by the following:
\begin{eqnarray}
&& {\cal M}^{\rm BdG}(2N; {\bf C}) = \left\{
{\cal H}=\left( \matrix{ a & b \cr b^{\dagger} & -a^{T} } \right): 
\right. \nonumber\\
&& \quad \mbox{$a$ is an $N \times N$ hermitian matrix,}
\nonumber\\
&& \quad \mbox{and
$b$ is an $N \times N$ complex symmetric matrix} \Big \}.
\nonumber
\end{eqnarray}
Since the dimension of the space of $a$ is $N^2$ 
and that of $b$ is $N(N+1)$,
the dimension of ${\cal M}^{\rm BdG}(2N; {\bf C})$ is
$N(2N+1)$.
Define
\begin{equation}
{\cal C}=\left( \matrix{0 & I_{N} \cr -I_{N} & 0} \right),
\label{eqn:C}
\end{equation}
where $I_{N}$ is the $N \times N$ unit matrix.
Then ${\cal H} \in {\cal M}^{\rm BdG}(2N; {\bf C})$ has the
following symmetry \cite{AZ96,AZ97}
\begin{equation}
{\cal H}=-{\cal C}{\cal H}^{T}{\cal C}^{-1},.
\label{eqn:sym}
\end{equation}
Assume that $\varphi_{j}$ is the $2N \times 1$ eigenvector of 
${\cal H}$ with an eigenvalue $\omega_{j}$;
$
   {\cal H} \varphi_{j} = \omega_{j} \varphi_{j}.
$
Then by (\ref{eqn:sym}),
$
-{\cal C} {\cal H}^{T} {\cal C}^{-1} \varphi_{j}
=\omega_{j} \varphi_{j}.
$
Take the complex conjugate of the both sides and
use the hermiticity of ${\cal H}$ and the fact 
${\cal C}^{-1}=-{\cal C}$, we have
$
   {\cal H} ({\cal C}\varphi_{j}^{*})
  =-\omega_{j} ({\cal C}\varphi_{j}^{*}).
$
That means ${\cal C}\varphi_{j}^{*}$ is the eigenvector of
${\cal H}$ with the eigenvalue $-\omega_{j}$.
Assume that $\omega_{1}, \omega_{2}, \cdots, \omega_{N}$
be the nonnegative eigenvalues of ${\cal H}$, then other 
eigenvalues are given by $-\omega_{1}, \cdots, -\omega_{N}$.
Therefore, if we set
\begin{equation}
\left( \matrix{U_{1}^{\dagger} \cr U_{2}^{\dagger}} \right)
\equiv (\varphi_{1}, \cdots, \varphi_{N}), \quad
U \equiv \left( \matrix{U_{1} & U_{2} \cr U_{2}^* & -U_{1}^* } \right),
\label{eqn:BdGU}
\end{equation}
then
$$
  {\cal H} U^{\dagger} = U^{\dagger} \Lambda, \quad
\mbox{with} \quad
\Lambda=\left(\matrix{\omega & 0 \cr 0 & - \omega} \right).
$$
We assume that $U$ is unitary.
Then we can see that $i U$ satisfies the relation
${\cal C} =(i U)^{T} {\cal C} (i U)$. 
The set of such $2N \times 2N$ unitary
matrices is called the symplectic group
${\rm Sp}(2N; {\bf C})$ \cite{FH91,Meh91}, whose dimension is
$\ell=2 N^2$.

The above consideration is summarized as follows; 
any ${\cal H} \in {\cal M}^{\rm BdG}(2N; {\bf C})$
can be diagonalized as
$$
U {\cal H} U^{\dagger}=\Lambda
=\left( \matrix{\omega & 0 \cr 0 & -\omega} \right), \quad
i U \in {\rm Sp}(2N; {\bf C}), 
$$
where $\omega={\rm diag}(\omega_{1}, \cdots, \omega_{N}), \,
\omega_{j} \geq 0, 1 \leq j \leq N$.
We then consider the map
\begin{equation}
{\cal H} \stackrel{\varphi}{\mapsto} (\omega, U)
=\Big((\omega_{j})_{1 \leq j \leq N},
{\bf p}=(p_{\mu})_{1 \leq \mu \leq \ell} \Big),
\label{eqn:map1}
\end{equation}
where ${\bf p}$ denotes the $\ell$-dimensional vector,
whose elements are the independent variables
of $U$. We have the Jacobian
of this map as \cite{AZ97},
\begin{eqnarray}
{\rm Jac}(\varphi) &=& \left|
{\rm det} \left(
\frac{\partial {\cal H}}{\partial \omega_{1}}, \cdots,
\frac{\partial {\cal H}}{\partial \omega_{N}},
\frac{\partial {\cal H}}{\partial p_{1}}, \cdots,
\frac{\partial {\cal H}}{\partial p_{\ell}} \right) \right|
\nonumber\\
&=& C({\bf p})
\prod_{j=1}^{N} \omega_{j}^2 
\prod_{1 \leq j < k \leq N}(\omega_{j}^2-\omega_{k}^2)^2,
\label{eqn:Jacobi1}
\end{eqnarray}
where $C({\bf p})$ is a function independent of
the eigenvalues $\omega$.

Next we consider the set
\begin{eqnarray}
&& {\cal M}^{\rm BdG}(2N; {\bf R}) = \left\{
{\cal H}=\left( \matrix{ a & b \cr b & -a^{T} } \right): \right.
\nonumber\\
&& \quad \left. 
\mbox{$a$ and $b$ are $N \times N$ real symmetric matrices} 
\right\}.
\nonumber
\end{eqnarray}
In this case, since the dimensions of the spaces of
$a$ and $b$ are both $N(N+1)/2$, the dimension of
${\cal M}^{\rm BdG}(2N; {\bf R})$ is $N(N+1)$.
We can see that 
any ${\cal H} \in {\cal M}^{\rm BdG}(2N; {\bf R})$
can be diagonalized as
$$
U' {\cal H} U'^{T}= \Lambda
=\left( \matrix{\omega & 0 \cr 0 & -\omega} \right), \quad
i U' \in {\rm Sp}(2N; i{\bf R}), 
$$
where $\omega={\rm diag}(\omega_{1}, \cdots, \omega_{N}), \,
\omega_{j} \geq 0, 1 \leq j \leq N$.
Here ${\rm Sp}(2N; i{\bf R})$ is the symplectic group
of $2N \times 2N$ matrices whose elements are
pure imaginary.
The map
\begin{equation}
{\cal H} \stackrel{\varphi'}{\mapsto} (\omega, U')
=\Big((\omega_{j})_{1 \leq j \leq N},
{\bf p}'=(p_{\mu})_{1 \leq \mu \leq \ell'} \Big),
\label{eqn:map2}
\end{equation}
with $\ell'=$dimension of ${\rm Sp}(2N; {\bf R})=N^2$,
is considered.
Here ${\bf p}'$ denotes the $\ell'$-dimensional vector
with the elements of the independent variables
of $U'$. The Jacobian
of this map is determined as \cite{AZ97}
\begin{equation}
{\rm Jac}(\varphi') 
= C'({\bf p}')
\prod_{j=1}^{N} |\omega_{j}|
\prod_{1 \leq j < k \leq N}|\omega_{j}^2-\omega_{k}^2|,
\label{eqn:Jacobi2}
\end{equation}
where $C'({\bf p}')$ is a function independent of
the eigenvalues $\omega$.
The important point is that ${\rm Jac}(\varphi)$
and ${\rm Jac}(\varphi')$ are proportional to
$|\h(\omega)|^2$ and $|\h(\omega)|$, respectively.

\subsection{Gaussian ensembles and eigenvalue 
distributions}

Altland and Zirnbauer introduced the Gaussian ensembles of the
BdG Hamiltonians in ${\cal M}^{\rm BdG}(2N; \C)$
and in ${\cal M}^{\rm BdG}(2N; \R)$, in which the
probability measures are in the form
\begin{equation}
\mu_{N}({\cal H}; \sigma^2) d{\cal H}
\propto \exp\left(-\frac{1}{4\sigma^2} 
{\rm tr} {\cal H}^2 \right) d{\cal H},
\label{eqn:meas1}
\end{equation}
with variance $2\sigma^2$.
For ${\cal H} \in {\cal M}^{\rm BdG}(2N; \C)$, 
we write the complex variables as
$a_{jk}=a_{jk}^{\rR}+i a_{jk}^{\rI},
b_{jk}=b_{jk}^{\rR}+i b_{jk}^{\rI}$ with
$i=\sqrt{-1}, a_{jk}^{\rR}, a_{jk}^{\rI},
b_{jk}^{\rR}, b_{jk}^{\rI} \in {\bf R}$, and choose
the independent variables as
$\{a_{jk}^{\rR}, b_{jk}^{\rR}, b_{jk}^{\rI}:
1 \leq j \leq k \leq N\} \cup
\{a_{jk}^{\rI}: 1 \leq j < k \leq N\}$.
Since
\begin{eqnarray}
&& {\rm tr} {\cal H}^2 =
\sum_{j, k} |{\cal H}_{jk}|^2 \nonumber\\
&=& 2 \sum_{j=1}^{N} \left\{ (a_{jj}^{\rR})^2
+(b_{jj}^{\rR})^2+(b_{jj}^{\rI})^2 \right\} \nonumber\\
&+& 4 \sum_{1 \leq j < k \leq N} \left\{
(a_{jk}^{\rR})^2+(a_{jk}^{\rI})^2
+(b_{jk}^{\rR})^2+(b_{jk}^{\rI})^2 \right\},
\nonumber
\end{eqnarray}
the probability measure (\ref{eqn:meas1}) is rewritten as
\begin{eqnarray}
&& \mu_{N}({\cal H}; \sigma^2) \nonumber\\
&=&
\prod_{j=1}^{N} 
\frac{e^{-(a_{jj}^{\rR})^2/2\sigma^2}}{\sqrt{2\pi \sigma^2}}
\frac{e^{-(b_{jj}^{\rR})^2/2\sigma^2}}{\sqrt{2\pi \sigma^2}}
\frac{e^{-(b_{jj}^{\rI})^2/2\sigma^2}}{\sqrt{2\pi \sigma^2}}
\nonumber\\
&\times& \prod_{1 \leq j < k \leq N} 
\frac{e^{-(a_{jk}^{\rR})^2/\sigma^2}}{\sqrt{\pi \sigma^2}}
\frac{e^{-(a_{jk}^{\rI})^2/\sigma^2}}{\sqrt{\pi \sigma^2}}
\nonumber\\
&\times& \prod_{1 \leq j < k \leq N} 
\frac{e^{-(b_{jk}^{\rR})^2/\sigma^2}}{\sqrt{\pi \sigma^2}}
\frac{e^{-(b_{jk}^{\rI})^2/\sigma^2}}{\sqrt{\pi \sigma^2}}
\label{eqn:meas2}
\end{eqnarray}
with the integration measure
\begin{equation}
d{\cal H}=\prod_{j=1}^{N} 
da_{jj}^{\rR}db_{jj}^{\rR} db_{jj}^{\rI}
\prod_{1 \leq j < k \leq N}
da_{jk}^{\rR} da_{jk}^{\rI} db_{jk}^{\rR} db_{jk}^{\rI}.
\label{eqn:meas3}
\end{equation}
The probability measure (\ref{eqn:meas1}) is given for
${\cal H} \in {\cal M}^{\rm BdG}(2N; \R)$,
by setting all
the imaginary parts $\{a_{jk}^{\rI}, b_{jk}^{\rI}\}$ be
zeros in (\ref{eqn:meas2}) and (\ref{eqn:meas3}).

On the other hand, following the maps (\ref{eqn:map1})
and (\ref{eqn:map2}), the integration measures are
transformed as
\begin{equation}
d{\cal H} \propto {\rm Jac}(\varphi) d \omega \times dU
\propto \h(\omega)^2 d \omega dU
\label{eqn:meas4}
\end{equation}
for ${\cal H} \in {\cal M}^{\rm BdG}(2N; \C)$,
$i U \in {\rm Sp}(2N; \C)$,
and
\begin{equation}
d{\cal H} \propto {\rm Jac}(\varphi') d \omega \times dU'
\propto |\h(\omega)| d \omega dU'
\label{eqn:meas5}
\end{equation}
for ${\cal H} \in {\cal M}^{\rm BdG}(2N; \R)$,
$iU' \in {\rm Sp}(2N; i \R)$,
respectively.
Since
${\rm tr}{\cal H}^2=2 |\omega|^2
=2 \sum_{j=1}^{N} \omega_{j}^2$, 
integrating over the spaces ${\rm Sp}(2N; \C)$ and
${\rm Sp}(2N; i \R)$ gives the distribution functions
of the nonnegative  
eigenvalues $\omega=(\omega_{1}, \cdots, \omega_{N}),
\omega_{i}\geq 0$, in the form of (\ref{eqn:BdG1})
with the indices $\alpha=\beta=2$
(class $C$) and with $\alpha=\beta=1$
(class $C$I), respectively.

\subsection{Harish-Chandra integral formula}

The transition probability density 
(\ref{eqn:g0}) from the state $\0$ 
to the state $\y$ in time $t$ is written as follows
using (\ref{eqn:calN}), where the proportional constants
are independent of the stochastic variables $\y$,
\begin{widetext}
\begin{eqnarray}
&& \gT(0, \0; t, \y)
\propto e^{-|\y|^2/2t} \h(\y) 
\int_{0 \leq z_{1} < \cdots < z_{N}} d\z \,
\f(T-t, \z|\y) \nonumber\\
&\propto& \h(\y) \int d\z \, {\rm sgn}
\left[ \h(\z) \right] 
\det_{1 \leq j, k \leq N} \left[
e^{-y_{j}^2/2t-(y_{j}-z_{k})^2/2(T-t)}
-e^{-y_{j}^2/2t-(y_{j}+z_{k})^2/2(T-t)} \right]
\nonumber\\
&=& \h(\y) \int d\z \, {\rm sgn}
\left[\h(\z)\right] e^{-|\z|^2/2T} \nonumber\\
&& \times \det_{1 \leq j, k \leq N} \left[
\exp\left(-\frac{T}{2t(T-t)} \left(
y_{j}-\frac{t}{T}z_{k} \right)^2 \right) 
- \exp \left(-\frac{T}{2t(T-t)} \left(
y_{j}+\frac{t}{T} z_{k} \right)^2 \right) \right].
\nonumber
\end{eqnarray}
Then we set $\omega'_{k}=tz_{k}/T, 1 \leq k \leq N$, and regard
$\omega'=(\omega'_{1}, \cdots, \omega'_{N})$ as the nonnegative 
eigenvalues of the BdG Hamiltonian 
${\cal H}'=({\cal H}'_{jk}) \in {\cal M}^{\rm BdG}(2N; \R)$.  
By (\ref{eqn:meas5}),
$ d \z \propto d \omega' \propto d{\cal H}'/|\h(\omega')|$, and we have
\begin{eqnarray}
&&\gT(0,\0; t, \y) \nonumber\\
&\propto& \h(\y) \int d{\cal H}' \,
\frac{1}{\h(\omega')} e^{-T|\omega'|^2/2t^2} 
\det_{1 \leq j,k \leq N} \left[
\exp \left(-\frac{T}{2t(T-t)}(y_{j}-\omega'_{k})^2 \right)
- \exp \left(-\frac{T}{2t(T-t)}(y_{j}+\omega'_{k})^2 \right)
\right]. \quad
\label{eqn:g2}
\end{eqnarray}
The result recently reported by Nagao \cite{Nag03} will
give a version of Harish-Chandra (Itzykson-Zuber)
integral formula in the present case,
\begin{eqnarray}
&&\int dU \, \exp \left(
-\frac{1}{4\sigma^2} {\rm tr}
(U^{\dagger} {\cal H} U-{\cal H}')^2 \right)
\propto \frac{1}{\h(\omega) \h(\omega')}
\det_{1 \leq j,k \leq N} \left[
e^{-(\omega_{j}-\omega'_{k})^2/2 \sigma^2}
-e^{-(\omega_{j}+\omega'_{k})^2/2 \sigma^2} \right],
\nonumber
\end{eqnarray}
where the integral is taken over the unitary matrices $U$
such that $i U \in {\rm Sp}(2N; \C)$, and
${\cal H}$ and ${\cal H}'$ are hermitian matrices in
${\cal M}^{\rm BdG}(2N; \C)$ and
${\cal M}^{\rm BdG}(2N; \R)$ having the
nonnegative eigenvalues $\omega=(\omega_{1}, \cdots, \omega_{N})$ and
$\omega'=(\omega'_{1}, \cdots, \omega'_{N})$, respectively.
Application of this identity to (\ref{eqn:g2}) gives
$$
\gT(0, \0; t, \y) \propto
\h(\y)^2 \int dU \int d {\cal H}'
\exp \left( -\frac{1}{2 (\sigma')^2} {\rm tr} ({\cal H}')^2 \right)
\exp \left( - \frac{1}{2 \sigma^2} {\rm tr}
(U^{\dagger} Y U - {\cal H}')^2 \right)
$$
\end{widetext}
with $\sigma^2=t(1-t/T)$ and $(\sigma')^2=t^2/T$, where
$Y={\rm diag}(y_{1}, \cdots, y_{N}, -y_{1}, \cdots, -y_{N})$,
$y_{j} \geq 0, 1 \leq j \leq N$.
This can be regarded as a BdG version of the two-matrix model
studied in \cite{KT02a} for the vicious walker model
without wall. 
Since it is a convolution of two Gaussian distributions, 
we will arrive at the equality
\begin{equation}
\gT(0, \0; t, \y) \propto
\h(\y)^2 \int dU 
\widehat{\mu}_{N,T}(t, U^{\dagger}Y U),
\label{eqn:g3}
\end{equation}
where, for ${\cal H} \in {\cal M}^{\rm BdG}(2N; \C)$,
\begin{eqnarray}
&& \widehat{\mu}_{N,T}(t, {\cal H}) \nonumber\\
&=&
\prod_{j=1}^{N} \left\{ 
\frac{e^{-(a_{jj}^{\rR})^2/2\sigma^2_{\rR}}}{\sqrt{2\pi \sigma^2_{\rR}}}
\frac{e^{-(b_{jj}^{\rR})^2/2\sigma^2_{\rR}}}{\sqrt{2\pi \sigma^2_{\rR}}}
\frac{e^{-(b_{jj}^{\rI})^2/2\sigma^2_{\rI}}}{\sqrt{2\pi \sigma^2_{\rI}}}
\right\} \nonumber\\
&\times& \prod_{1 \leq j < k \leq N} \left\{
\frac{e^{-(a_{jk}^{\rR})^2/\sigma^2_{\rR}}}{\sqrt{\pi \sigma^2_{\rR}}}
\frac{e^{-(a_{jk}^{\rI})^2/\sigma^2_{\rI}}}{\sqrt{\pi \sigma^2_{\rI}}}
\right\} \nonumber\\
&\times& \prod_{1 \leq j < k \leq N} \left\{
\frac{e^{-(b_{jk}^{\rR})^2/\sigma^2_{\rR}}}{\sqrt{\pi \sigma^2_{\rR}}}
\frac{e^{-(b_{jk}^{\rI})^2/\sigma^2_{\rI}}}{\sqrt{\pi \sigma^2_{\rI}}}
\right\}
\label{eqn:hatmu}
\end{eqnarray}
with
$$
\sigma^2_{\rR}=\sigma^2+(\sigma')^2=t, \quad
\sigma^2_{\rI}=\sigma^2=t \left( 1-\frac{t}{T} \right).
$$
Now the transition from the class $C$ to the class $C$I
is explicitly represented.
The variance $\sigma^2_{\rR}$ increases linearly in $t$,
but $\sigma^{2}_{\rI}$ increases in time only up to
time $t=T/2$ and then decreases in time.
As $t \to T$, $\sigma^2_{\rI} \to 0$
making the imaginary parts of matrix elements zeros
with probability one, and the symmetry class 
is changed.

\section{MOMENTS OF VICIOUS WALKERS WITH A WALL}

In this section, we study the moments of positions
of vicious walkers with a wall in order to characterize
the transition of distribution, as reported in
\cite{KKom03} for the vicious walkers without wall.
The $n$-th moment of the positions of walkers is
defined as
\begin{eqnarray}
&& m_{N,T}(t, n) = \left\langle \sum_{j=1}^{N} x_{j}^{n}
\right\rangle_{t} \nonumber\\
&& \quad = \int_{0 \leq x_{1} < \cdots < x_{N}} d \x \
\sum_{j=1}^{N} x_{j}^{n} 
\gT(0, \0; t, \x)
\nonumber
\end{eqnarray}
for $n=1,2, \cdots$, where $x_{j}$ denotes the 
position of $j$-th walker.

\subsection{Wick's formula}

By (\ref{eqn:meas4}) and the equality (\ref{eqn:g3}),
if $n=2k$ is even, $k=1,2, \cdots$, we have the equality
\begin{equation}
m_{N,T}(t, 2k)= \frac{1}{2}
\left\langle {\rm tr} {\cal H}^{2k} \right\rangle,
\label{eqn:Mk2}
\end{equation}
where 
$$
\left\langle {\rm tr} {\cal H}^{2k} \right\rangle
= \int {\rm tr} {\cal H}^{2k} 
\widehat{\mu}_{N,T}(t, {\cal H}) d {\cal H}.
$$
Note that
$$
  {\rm tr}({\cal H}^{2k})=\sum_{j_{1}, j_{2}, \cdots, j_{2k}}
   {\cal H}_{j_{1} j_{2}} {\cal H}_{j_{2} j_{3}} \cdots
{\cal H}_{j_{2k-1} j_{2k}} {\cal H}_{j_{2k} j_{1}},
$$
where the sum is taken over all $N^{2k}$ combinations
of indices $j_{1}, j_{2}, \cdots , j_{2k}$.

Since (\ref{eqn:hatmu}) is a product of independent
Gaussian integration-kernels, we can apply the
Wick formula with the variances
\begin{eqnarray}
&& \langle (a_{j \ell}^{\rR})^2 \rangle = 
\frac{\sigma_{\rR}^2}{2}(1+\delta_{j \ell}), \quad
\langle (a_{j \ell}^{\rI})^2 \rangle = 
\frac{\sigma_{\rI}^2}{2}(1-\delta_{j \ell}), \nonumber\\
&& \langle (b_{j \ell}^{\rR})^2 \rangle = 
\frac{\sigma_{\rR}^2}{2}(1+\delta_{j \ell}), \quad
\langle (b_{j \ell}^{\rI})^2 \rangle = 
\frac{\sigma_{\rI}^2}{2}(1+\delta_{j \ell}), \qquad
\nonumber
\end{eqnarray}
for $1 \leq j \leq \ell \leq N$, 
where $\delta_{j \ell}$ is 
Kronecker's delta.
These relations are rewritten as 
\begin{eqnarray}
&& \langle a_{j \ell} a_{m n} \rangle
=  \left\langle a_{j \ell} (a^{T})_{n m} \right\rangle
= \frac{c^2}{2} (\delta_{j n } \delta_{\ell m}
+ \gamma \delta_{j m} \delta_{\ell n} ), \nonumber\\
&& \langle b_{j \ell} b_{mn} \rangle
= \frac{c^2}{2} \gamma 
(\delta_{j n} \delta_{\ell m}
+ \delta_{j m} \delta_{\ell n} ), \nonumber\\
&& \left\langle b_{j \ell} (b^{\dagger})_{m n} \right\rangle
= \frac{c^2}{2} 
(\delta_{j n} \delta_{\ell m}
+ \delta_{j m} \delta_{\ell n} ),
\nonumber
\end{eqnarray}
for 
$1 \leq j, \ell, m, n \leq N$, where
$$
c^2 = \frac{t(2T-t)}{T}, \quad
\gamma = \frac{t}{2T-t}.
$$
Define
\begin{eqnarray}
&& \delta_{N}(j, \ell; m, n) =
\delta_{jn} \delta_{\ell m} 
- \delta_{j+N \, m} \delta_{\ell+N \, n} 
- \delta_{j-N \, m} \delta_{\ell-N \, n} 
\nonumber\\
&& \qquad
+ \delta_{j+N \, m} \delta_{\ell-N \, n}
+ \delta_{j-N \, m} \delta_{\ell+N \, n}.
\label{eqn:bracket0}
\end{eqnarray}
Then we have the variance of the BdG-type Hamiltonian
in our time-dependent ensemble as
\begin{equation}
\langle {\cal H}_{j \ell} {\cal H}_{m n} \rangle
= \frac{c^2}{2} \Big\{
\delta_{N}(j, \ell; m, n)+
\gamma \delta_{N}(j, \ell; n, m) \Big\}
\label{eqn:Wick3}
\end{equation}
for $1 \leq j, \ell, m, n \leq 2N$.
The Wick formula for (\ref{eqn:Mk2}) is thus
\begin{widetext}
\begin{eqnarray}
m_{N,T}(t, 2k) 
&=& \frac{1}{2} \sum_{j_{1}, j_{2}, \cdots, j_{2k}}
\sum_{\pi \in S_{2k}:{\rm R}}
\langle {\cal H}_{j_{\pi(1)} j_{\pi(1)+1}} {\cal H}_{j_{\pi(2)} j_{\pi(2)+1}} 
\rangle
\langle {\cal H}_{j_{\pi(3)} j_{\pi(3)+1}} {\cal H}_{j_{\pi(4)} j_{\pi(4)+1}} 
\rangle \nonumber\\
&& \hskip 5cm
\cdots
\langle {\cal H}_{j_{\pi(2k-1)} j_{\pi(2k-1)+1}} 
{\cal H}_{j_{\pi(2k)} j_{\pi(2k)+1}}
\rangle,
\label{eqn:Mk4}
\end{eqnarray}
\end{widetext}
with the identification $j_{2k+1}=j_{1}$, 
where the first sum is taken over all $N^{2k}$ combinations
of indices $j_{1}, j_{2}, \cdots, j_{2k}$,
and the second one over the set of permutations $S_{2k}$ 
of $\{1,2, \cdots, 2k\}$ with the restriction
\begin{eqnarray}
{\rm R}: &&
\pi(1) < \pi(3) < \cdots < \pi(2k-1), \nonumber\\
&& \pi(2j-1) < \pi(2j), \quad 1 \leq j \leq k.
\nonumber
\end{eqnarray}
The total number of the terms in the second summation is
$(2k-1)!!$.

\subsection{M\"obius graph expansion}

There are ten terms in the variance (\ref{eqn:Wick3}) with
(\ref{eqn:bracket0}). We will represent each of them
by a pair of lines as shown in Fig.3,
by expressing Kronecker's delta $\delta_{jn}$
by a line without arrow connecting $j$ and $n$,
$\delta_{j+N \, m}$ by a line with arrow
in the direction from $j$ to $m$,
and $\delta_{j-N \, m}$ by a line with arrow
in the direction from $m$ to $j$, respectively.
The weights, which should be multiplied to the
factor $c^2/2$, are also listed.
We regard these pairs of lines as the hems of ribbons.
As shown in Fig.3, in ten kinds of ribbons,
half of them are twisted and others are untwisted.
Two kinds of ribbons do not have any arrows on hems, and
other eight kinds of ribbons have arrows on hems.
We will call the ribbons having arrows
in the same direction {\it current ribbons}
(c-ribbons for short), the ribbons having 
arrows in the opposite directions {\it exchange ribbons}
(e-ribbons), and those not having any arrows
{\it normal ribbons} (n-ribbons), respectively.

\begin{figure}
\includegraphics[width=.8\linewidth]{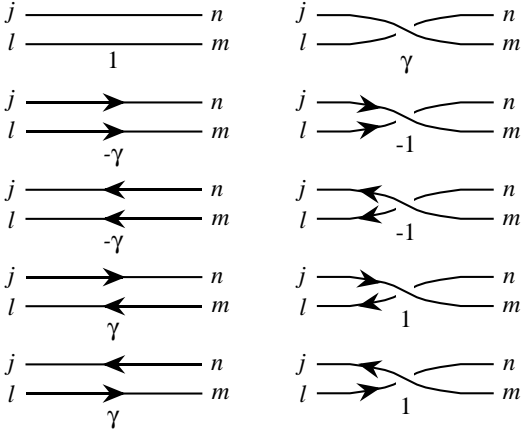}
\caption{Ten kinds of ribbons with and without
arrows. Weights are also listed. \label{fig:fig3}}
\end{figure}

Inserting (\ref{eqn:Wick3}) into (\ref{eqn:Mk4}) gives
a sum of the $K=(2k-1)!! \times 10^{k}$
terms in the form,
$m_{N,T}(t, 2k)=(1/2) \cdot (c^2/2)^k
\sum_{\ell=1}^{K} L_{\ell}$ with
$L_{\ell}=\sum_{j_{1}, \cdots, j_{2k}} L_{\ell}
(j_{1}, \cdots, j_{2k})$.
As explained in \cite{KKom03}, each term 
$L_{\ell}(j_{1}, \cdots, j_{2k})$ is represented by
a graph which consists of a $2k$-gon with its edges
$\overline{j_{1} j_{2}}, \overline{j_{2} j_{3}},
\cdots, \overline{j_{2k} j_{1}}$ connected by
$k$ ribbons to make $k$ pairs (Wick pairs). Such graphs may be called
{\it M\"obius graphs having ribbons with and without arrows}.
If we assume that there are $k_{n}$ n-ribbons,
$k_{c}$ c-ribbons and $k_{e}$ e-ribbons,
and among them
$\varphi_{n}$ n-ribbons, $\varphi_{c}$ c-ribbons
and $\varphi_{e}$ e-ribbons are twisted, the weight
of $L_{\ell}(j_{1}, \cdots, j_{2k})$ is
$\gamma^{\varphi_{n}} \times (-\gamma)^{k_{c}-\varphi_{c}}
\times (-1)^{\varphi_{c}} \times \gamma^{k_{e}-\varphi_{e}}
=(-1)^{k_{c}} \gamma^{k-k_{n}-\varphi+2 \varphi_{n}}$,
where $k=k_{n}+k_{c}+k_{e}$ and
$\varphi \equiv \varphi_{n}+\varphi_{c}+ \varphi_{e}$
(the total number of twisted ribbons).
For each vertex $j_{s}, 1 \leq s \leq 2k$, we take the
summation of the index over $1 \leq j_{s} \leq 2N$
to calculate $L_{\ell}$ from
$\{L_{\ell}(j_{1}, \cdots, j_{2k})\}$.
Since each ribbon represents a product of two
Kronecker's deltas in (\ref{eqn:bracket0}),
any pairs of indices $j_{s}$ and $j_{s}'$ connected by
a line (a hem of ribbon) should be identified,
or identified in modulus $\pm N$,
and the free indices remaining after this
``identification" of indices give $N$ dependence
to $L_{\ell}$. We will find the following rules
for the $N$ dependence,
where it should be noted that each vertex 
is the endpoint of two lines (two hems of two ribbons)
with or without arrows.
(i) If both of the lines connected to a vertex have no
arrows, then we will call such a vertex a {\it 0-vertex}.
The summation over a free index on a 0-vertex gives
$2N$. 
(ii) If at least one of the two lines connected to a
vertex has an arrow and they are not a pair of lines
with inward and outward arrows,
then the summation over a free index
on such a vertex gives $N$.
(iii) Otherwise, the sum over free index becomes zero.
Consider the equivalence classes of the M\"obius graphs
with n-, c-, and e-ribbons. 
If a class is expressed by a representative graph,
say $\Gamma$, the number of elements of the class $\Gamma$
({\it i.e.} the number of graphs topologically
equivalent with $\Gamma$) is denoted by $|\Gamma|$.
Let $V(\Gamma)$  be the total numbers
of free indices of $\Gamma$ and $V_{0}(\Gamma)$ be the
number of free indices on 0-vertices.
Set $\widehat{\cal G}(k)$ be the collection of all such graphs
$\{\Gamma\}$, then we have the combinatorial expression
for the moments as
\begin{eqnarray}
&& m_{N,T}(t, 2k) = 
\frac{1}{2}\left(\frac{c^2}{2}\right)^k
\sum_{\Gamma \in \widehat{\cal G}(k)} |\Gamma|
2^{V_{0}(\Gamma)} N^{V(\Gamma)} \nonumber\\
&& \qquad \times
(-1)^{k_{c}(\Gamma)}
\gamma^{k-k_{n}(\Gamma)-\varphi(\Gamma)
+2\varphi_{n}(\Gamma)}.
\nonumber
\end{eqnarray}

\begin{figure}
\includegraphics[width=.7\linewidth]{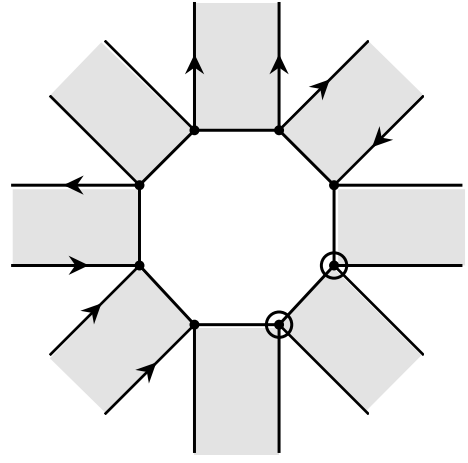}
\caption{An example of allowed ways of
putting arrows on ribbons. 
The 0-vertices are marked by circles.\label{fig:fig4}}
\end{figure}

In \cite{KKom03} the collection ${\cal G}$ of
topologically distinct M\"obius graphs, which consist
of $2k$-gons and $k$ normal ribbons, was introduced
and the fact was used that
each graph $\Gamma \in {\cal G}$ having only untwisted
ribbons (resp. having some twisted ribbons)
defines an orientable (resp. non-orientable) surface
$S_{\Gamma}$ by a {\it map} \cite{Zvo97,Oko99} 
to derive the $1/N^2$ (resp.$1/N$) expansion.
The number of distinct orientable (resp. non-orientable)
surfaces with genus $g$ obtained from the graphs
without any twisted ribbons (resp. with $m$ twisted ribbons) in
${\cal G}$ was denoted by $\varepsilon_{g}(k)$ \cite{HZ86}
(resp. $\widetilde{\varepsilon}_{g,m}(k)$ \cite{KKom03}).
We notice that the M\"obius graphs in 
$\widehat{\cal G}(k)$ introduced here are obtained 
by putting arrows on hems of some of the normal ribbons
in $\Gamma \in {\cal G}(k)$.
There only allowed ways to put them are such that
there is no pairs of lines with inward and 
outward arrows connected to a vertex (see Fig.4). 
We now introduce the following multiplicative
factors to $\varepsilon_{g}(k)$ and
$\widetilde{\varepsilon}_{g,m}(k)$,
$A_{v_{0}, \kappa_{n}, \kappa_{c}}^{g}(k)
\equiv$ the number of allowed ways to put arrows 
on the hems of ribbons in a M\"obius graph 
without twisted ribbons, which is mapped to a
surface with genus $g$,
so that the graph becomes to have 
$\kappa_{n}$ n-ribbons, $\kappa_{c}$ c-ribbons,
and $v_{0}$ free indices on 0-vertices, and
$\widetilde{A}_{v_{0}, \kappa_{n}, \kappa_{c}, m_{n}}^{g,m}(k)
\equiv$ the number of allowed ways to put arrows 
on the hems of ribbons in a M\"obius graph 
with $m$ twisted ribbons, which is mapped to a
surface with genus $g$,
so that the graph becomes to
have $\kappa_{n}$ n-ribbons in which $m_{n}$ are twisted,
$\kappa_{c}$ c-ribbons,
and $v_{0}$ free indices on 0-vertices.
Here remark that the total number of ribbons is
fixed to be $k$,
the number of e-ribbons should be
$k-\kappa_{n}-\kappa_{c}$, and that the total number of free
vertices $V(\Gamma)$ is determined by $k$ and $g$ through
the relations with the Euler characteristics $\chi \equiv V-k+1$ as
$\chi=2-2g$ for $m=0$ and
$\chi=2-g$ for $m \geq 1$, respectively.
Then we have the following $1/N$ expansion formula
for the moments,
\begin{eqnarray}
&& m_{N,T}(t, 2k) = \frac{1}{2} \left(\frac{c^2}{2} \right)^k
N^{k+1} \sum_{g=0}^{[k/2]} \varepsilon_{g}(k)
\left(\frac{1}{N^2}\right)^{g} \nonumber\\
&\times& \sum_{v_{0}=0}^{k+1-2g} 
\sum_{\kappa_{n}=0}^{k} \sum_{\kappa_{c}} (-1)^{\kappa_{c}}
2^{v_{0}}  A_{v_{0}, \kappa_{n}, \kappa_{c}}^{g}(k)
\gamma^{k-\kappa_{n}}
\nonumber\\
&+& \frac{1}{2} \left(\frac{c^2}{2} \right)^k
N^{k+1} \sum_{g=1}^{k} \left(\frac{1}{N}\right)^{g}
\sum_{m=1}^{k} \widetilde{\varepsilon}_{g,m}(k)
\sum_{v_{0}=0}^{k+1-g} 2^{v_{0}} 
\nonumber\\
&\times& 
\sum_{\kappa_{n}=0}^{k} \sum_{\kappa_{c}}
\sum_{m_{n}=0}^{m} (-1)^{\kappa_{c}} 
\widetilde{A}_{v_{0}, \kappa_{n}, \kappa_{c}, m_{n}}^{g,m}(k)
\gamma^{k-\kappa_{n}-m+2m_{n}}. \nonumber\\
\label{eqn:combi2}
\end{eqnarray}
Here we may prove that
$\widetilde{A}_{v_{0}, \kappa_{n}, \kappa_{c}, m_{n}}^{g,m}(k)
=0$ for $k-\kappa_{n}-m+2m_{n} <0$, and
(\ref{eqn:combi2}) will give a series with nonnegative powers
of $\gamma$.

As an example, we consider the fourth moment.
For $k=2$, we found \cite{KKom03}
\begin{eqnarray}
&& \varepsilon_{0}(2)=2, \quad
\varepsilon_{1}(2)=1, \nonumber\\
&& \widetilde{\varepsilon}_{1,1}(2)=4, \quad
\widetilde{\varepsilon}_{1,2}(2)=1, \quad
\widetilde{\varepsilon}_{2,1}(2)=2, \quad
\widetilde{\varepsilon}_{2,2}(2)=2.
\nonumber
\end{eqnarray}
As we can confirm easily that the factors 
$A_{v_{0}, \kappa_{n}, \kappa_{c}}^{g}(2)$
and $\widetilde{A}_{v_{0}, \kappa_{n}, \kappa_{c}, m_{n}}^{g,m}(2)$
have non-zero values only in the following cases,
\begin{eqnarray}
&& A_{3,2,0}^{0}(2)=1, \quad
A_{1,2,0}^{1}(2)=1, \nonumber\\
&& A_{0,1,1}^{1}(2)=4, \quad
A_{0,0,0}^{1}(2)=2, \nonumber\\
&& \widetilde{A}_{1,1,0,0}^{1,1}(2)=2, \quad
\widetilde{A}_{2,2,0,1}^{1,1}(2)=1, \quad
\widetilde{A}_{0,0,1,0}^{1,2}(2)=4, \nonumber\\
&& \widetilde{A}_{2,2,0,2}^{1,2}(2)=1, \quad
\widetilde{A}_{0,1,1,0}^{2,1}(2)=2, \quad
\widetilde{A}_{1,2,0,1}^{2,1}(2)=1, \nonumber\\
&& \widetilde{A}_{0,0,1,0}^{2,1}(2)=2, \quad
\widetilde{A}_{0,1,0,1}^{2,1}(2)=2, \quad
\widetilde{A}_{0,0,0,0}^{2,2}(2)=2, \nonumber\\
&& \widetilde{A}_{0,1,0,1}^{2,2}(2)=4, \quad
\widetilde{A}_{1,2,0,2}^{2,2}(2)=1. 
\nonumber
\end{eqnarray}
Then (\ref{eqn:combi2}) gives
\begin{eqnarray}
&& m_{N,T}(t,4) 
=\frac{c^{4}}{4}
\Big\{ 8 N^3 + (6 + 8 \gamma+2 \gamma^2) N^2 \nonumber\\
&& \qquad \qquad 
+(1+2 \gamma+5 \gamma^2) N \Big\}.
\label{eqn:fourth}
\end{eqnarray}

By definition we will see
$A_{v_{0},\kappa_{n}, \kappa_{c}}^{0}(k)=
\delta_{v_{0} \, k+1} \delta_{\kappa_{n} k}
\delta_{\kappa_{c} 0}$, and thus
the leading term in (\ref{eqn:combi2}) for
large $N$ is
$$
m_{N,T}(t,2k) =
\frac{1}{2} \left(\frac{c^2}{2}\right)^k
(2N)^{k+1} \varepsilon_{0}(k) 
+{\cal O}(N^k).
$$
Since $\varepsilon_{0}(k)$ is the Catalan number,
Wigner's semicircle law will hold in $N \to \infty$
also in the BdG random matrices.

\subsection{Calculation by density function}

By (\ref{eqn:g0}) with (\ref{eqn:calN}) and (\ref{eqn:hath}),
it is easy to see that $\gT(0, \0; t, \x)$ is symmetric in
$x_{1}, \cdots, x_{N}$. 
Then we can define the density function as
$$
\widehat{\rho}(t,\x) = \frac{1}{(N-1)!} \int_{0}^{\infty}
\prod_{j=2}^{N} dx_{j} \gT(0, \0; t, \x)
$$
and the even moments (\ref{eqn:Mk2}) are calculated by it as
\begin{equation}
m_{N,T}(t, 2k) = \int_{0}^{\infty} x^{2k}
\widehat{\rho}(t,x) dx.
\label{eqn:Mden1}
\end{equation}

Let $L_{j}^{(a)}(z)$ be the Laguerre polynomials
with a parameter $a$ defined as
$$
L_{j}^{(a)}(z) = \frac{e^z z^{-a}}{j !} 
\frac{d^{j}}{dz^{j}} (e^{-z} z^{j+a}).
$$
Using them with $a=1/2$, we can define the monic polynomials
$
C_{j}(z)=(-1)^{j} j ! L_{j}^{(1/2)}(z),
$
which satisfy the orthogonality
$
\int_{0}^{\infty} z^{1/2} e^{-z} C_{j}(z) C_{\ell}(z) dz
= h_{j} \delta_{j \ell}
$
with $h_{j}=\Gamma(j+3/2) j !$. 
Quite recently Nagao gave the general expressions of
dynamical correlations for the present system
using $C_{j}(z)$
\cite{Nag03}. From his result, we can read the
density function $\widehat{\rho}(t, x)$ as
\begin{widetext}
\begin{eqnarray}
&& \widehat{\rho}(t, x) =
\frac{2x^2}{c^3} e^{-(x/c)^2} \sum_{j=0}^{N-1}
\frac{C_{j}\left((x/c)^2\right)^2}{h_{j}}
+ \frac{2x^2}{c^3} e^{-(x/c)^2}
\sum_{j=N}^{\infty} \sum_{\ell=0}^{N-1} \sum_{m=0}^{\ell}
\beta_{j \ell} \alpha_{\ell m} 
\frac{C_{j}\left( (x/c)^2 \right)
C_{m} \left( (x/c)^2 \right)}{h_{j}}
\gamma^{j-m},
\nonumber
\end{eqnarray}
where
\begin{eqnarray}
\alpha_{2j \, \ell} &=& (-1)^{\ell} \frac{(2j)!}{\sqrt{\pi}}
\frac{\Gamma(2j-\ell+1/2)}{(2j-\ell)! \ell!}, \quad
\alpha_{2j+1 \, \ell} = (-1)^{\ell}
\frac{(2j+1)!}{\sqrt{\pi}}
\left(2j-\ell+\frac{1}{4} \right) 
\frac{\Gamma(2j-\ell-1/2)}{(2j-\ell+1)! \ell!}, \nonumber\\
\beta_{j \, 2\ell} &=& \frac{(-1)^{j+1}}{2 \sqrt{\pi}}
\frac{\Gamma(j-2\ell-1/2)}{(j-2\ell)!} \frac{j!}{(2\ell)!}, \quad
\beta_{j \, 2\ell+1} = \frac{(-1)^j}{2 \sqrt{\pi}}
\frac{j!}{(2\ell+1)!} \sum_{m=0}^{[(j-2\ell-1)/2]}
\frac{\Gamma(j-2\ell-2 m-3/2)}{(j-2\ell-2 m-1)!}.
\nonumber
\end{eqnarray}

Let $H_{j}(z)$ be the $j$-th Hermite polynomial,
$H_{j}(z)=(-1)^{j} e^{z^2} (d/dz)^{j} e^{-z^2}$.
If we use the following equalities \cite{Bat53,Bat54},
\begin{eqnarray}
&& L_{j}^{(a-1)}(z) = L_{j}^{(a)}(z)-L_{j-1}^{(a)}(z), 
\quad
z L_{j}^{(a+1)}(z) = (j+a+1) L_{j}^{(a)}(z)
-(j+1) L_{j+1}^{(a)}(z), \nonumber\\
&& \sum_{m=0}^{j} \frac{m!}{\Gamma(m+a+1)} (L_{m}^{(a)}(z))^2
= \frac{(j+1)!}{\Gamma(j+a+1)}
\left\{ L_{j}^{(a)}(z) L_{j}^{(a+1)}(z)
-L_{j+1}^{(a)}(z) L_{j-1}^{(a+1)}(z) \right\}, \nonumber\\
&& L_{j}^{(-1/2)}(z) = \frac{(-1)^j}{2^{2j}j!} H_{2j}(\sqrt{z}),
\quad
\sqrt{z} L_{j}^{(1/2)}(z) = \frac{(-1)^{j}}{2^{2j+1} j!}
H_{2j+1}(\sqrt{z})
\quad \mbox{for} \ z \geq 0,
\nonumber
\end{eqnarray}
we will have the following expression for
the density function, 
\begin{eqnarray}
&& \widehat{\rho}(t, x) = 
\frac{N}{2^{4N-3} (N-1)! \Gamma(N+1/2)} \frac{1}{c} e^{-(x/c)^2}
\nonumber\\
&& \times \left\{
(H_{2N-1}(x/c))^2-\frac{(2N-1)(N-1)}{N} \frac{c}{x}
H_{2N}(x/c) H_{2N-3}(x/c) 
-\frac{N-1}{2N} \frac{c}{x}
H_{2N}(x/c) H_{2N-1}(x/c) \right\} \nonumber\\
&& + \frac{1}{c} \sum_{j=N}^{\infty}
\sum_{\ell=0}^{N-1}
\sum_{m=0}^{\ell} 
\beta_{j \ell} \alpha_{\ell m}
\gamma^{j-m} 
\frac{1}{2^{2m} (2j+1)! \sqrt{\pi}}
e^{-(x/c)^2} 
H_{2j+1}(x/c) H_{2m+1}(x/c). 
\label{eqn:rhob2}
\end{eqnarray}
Substituting (\ref{eqn:rhob2}) into (\ref{eqn:Mden1}) and 
using the integral formulae of Hermite polynomials used in
\cite{KKom03}
and the relation
$\sum_{\ell=m}^{n} \beta_{n \ell} \alpha_{\ell m}
=\delta_{n m}$ for $n \geq m$ \cite{Nag03}, 
we will arrive at the expression
\begin{eqnarray}
&&m_{N,T}(t, 2k) =\left(\frac{c^2}{2}\right)^{k} 
N (2k-1)!!
\sum_{j=0}^{k}{2N-1 \choose k-j} 2^{k-j} 
\left[ {k \choose j} 
-{k-1 \choose j} \left\{
\frac{k-j-1}{2(k-j+1)}+\frac{N-1}{2N-k+j} \right\}
\right] \nonumber\\
&& - \left(\frac{c^2}{2}\right)^{k} 
\sum_{j=0}^{k-1} \sum_{\ell=0}^{N-1} 
\sum_{m=\ell}^{k-j-1} \gamma^{m+1}
\sum_{n=N}^{m-\ell+N} 
\beta_{m-\ell+N \, n} \alpha_{n \, N-\ell-1} 
\nonumber\\
&& \hskip 5cm \times
\frac{(2k)! (2N-2\ell-1)! 2^{m-j+1}}
{j! (j+2N-2\ell+m-k)! (k-j-m-1)!(k-j+m+1)!}.
\label{eqn:Mden2}
\end{eqnarray}
\end{widetext}
As a matter of course, when we set $k=2$, (\ref{eqn:Mden2})
gives the fourth moment (\ref{eqn:fourth}).
The large-$N$ behavior discussed below (\ref{eqn:fourth})
is obtained also from (\ref{eqn:Mden2}).

\section{CONCLUDING REMARKS}

In the present paper we have considered the solvability
of one-dimensional vicious walker models, which are
generalizations of the model studied in earlier
papers \cite{KT02a,NKT03}.
We are interested in these systems as the nonequilibrium
statistical models, since they provide in general spatially
and temporally inhomogeneous systems.
The temporally inhomogeneous one-particle
systems have been extensively studied by Yor
by constructing them from Brownian motions
and Bessel processes \cite{Yor92}.
Our present work may be regarded as a new attempt to
construct many-particle systems in one-dimension
using such temporally inhomogeneous processes as
elementary processes, so that we can discuss also the
spatial  inhomogeneity.

We have imposed the noncolliding condition between
particles, since we have thought them as
{\it vicious walkers}. This condition introduces
the effective long-ranged repulsive interactions
among particles and it may make the models be
nontrivial many-particle systems.
Our strategy to analyze the models is to map these
interacting particle systems to multimatrix models
defined in the appropriate matrix spaces, which have
in general higher dimensions than the original
phase spaces of particle systems.
The particle positions are expressed by the
statistics of eigenvalues \cite{Dys62}
or the ``radial coordinates" \cite{Hua63,VZ93,Ver94,JSV96,SV98}
of random matrices.

We have proved the following equivalences in probability
distributions.
\begin{description}
\item{(i)} \quad The noncolliding system of 
Brownian motions with a wall,
defined as the continuum limit of the 
vicious walker model with a wall,
and the noncolliding system of Brownian meanders,
which are constructed from the three-dimensional
Bessel process.
\item{(ii)} \quad
The noncolliding $N$ generalized meanders
constructed from the $d$-dimensional Bessel processes
with $d=2,4,6, \cdots$ and the
``radial coordinate" of the
$\{N+(d-2)/2\} \times N$ rectangular random matrices
in the Gaussian ensemble with time-dependent
variances (a two-matrix model of
the chiral GUE and chiral GOE.)
\item{(iii)} \quad
The noncolliding $N$ Brownian meanders
and the nonnegative eigenvalues of $2N \times 2N$
BdG-type random Hamiltonians in the Gaussian
ensemble with time-dependent variances
(a two-matrix model of the BdG-type random matrices
in the classes $C$ and $C$I of Altland and Zirnbauer).
\end{description}
We have discussed the noncolliding system of generalized
meanders made from the general odd-dimensional Bessel processes.
So a natural question is what is the corresponding
random matrix theory for the noncolliding system
of generalized meanders associated with the $d$-dimensional
Bessel processes with $d=5,7,9, \cdots$.
As far as we know, it is an open problem.

We have claimed that the above equivalence between the
interacting particle systems and the Gaussian multimatrix
models implies the solvability of the systems.
This statement may be true, but to obtain exact expressions
of general correlation functions for the 
multimatrix models \cite{NF99,Nag01}
is far from trivial and one
have to use a series of techniques developed in
the random matrix theory.
Exact expressions of general dynamical correlation functions
enable us to discuss the infinite particle limit $N \to \infty$
of the nonequilibrium system as reported in
\cite{NKT03,KNT03} for the vicious walker model
(noncolliding Brownian motions) without wall, 
and in \cite{Nag03} for the system with a wall.
We expect that the infinite particle limits of dynamical 
correlations for the noncolliding generalized meanders 
can be generally evaluated by following the strategy 
employed in \cite{Nag03,FNH99}.

As a simplest case of the correlation functions,
the density function can be determined.
As reported in \cite{KKom03}, calculation of 
the moments of the positions of vicious walkers
is related with an enumeration problem
of orientable and non-orientable surfaces with a
fixed number of genus, which are obtained from
the M\"obius graphs with a fixed number of
twisted ribbons by {\it map} \cite{Zvo97,Oko99}.
In the present paper, we showed that a new graphical
problem arises from the vicious walker model with 
wall restriction; an enumeration problem of the
ways of assigning arrows on the ribbons of 
M\"obius graphs following some rules.
It should be noted that, roughly speaking,
there were two kinds of
ribbons in our expansion formula, with and without
arrows. These ribbons are thinned into lines and
they are drawn on surfaces.
Then this enumeration problem may provide, if
we consider the lines as ``world lines" of particles, 
the statistical mechanics of composite particles
on random surfaces.

Recently a variety of problems associated with the
conditional random walks/Brownian motions have been
proposed and intensively studied in statistical physics,
{\it e.g.} first passage problem \cite{Red01},
``lion-lamb" problem \cite{KR96,RK99}, 
diffusion particle systems with mobile traps \cite{BB02},
families of vicious walkers \cite{CK03},
leader and laggard problem \cite{AJMKR02},
system of stochastic Loewner evolutions \cite{Car03},
friendly walker models \cite{TK98,CC99,GV02,IK01,TY03}
and so on. 
Solvability and unsolvability of these models
will be important topics in statistical physics
far from equilibrium.

\vskip 1cm

\begin{acknowledgments}
MK thanks J. Cardy,
P. J. Forrester and T. Fukui for useful comments at the
very beginning of the present work.
MK and HT acknowledge the valuable communication 
with M. Yor.
\end{acknowledgments}


\begin{thebibliography}{99}
\bibitem{Lig85}
T. M. Liggett, 
{\it Interacting Particle Systems},
(Springer, New York, 1985).

\bibitem{KKon93}
M. Katori and N. Konno,
J. Phys. A {\bf 26}, 6597 (1993).

\bibitem{Lig99}
T. M. Liggett,
{\it Stochastic Interacting Systems: Contact, Voter,
and Exclusion Processes},
(Springer, New York, 1999).

\bibitem{DEHP93}
B. Derrida, M. R. Evans, V. Hakim and V. Pasquier,
J. Phys. A {\bf 26}, 1493 (1993).

\bibitem{Sas99}
T. Sasamoto, 
J. Phys. A {\bf 32}, 7109 (1999);
J. Phys. Soc. Jpn. {\bf 69}, 1055 (2000).

\bibitem{Fis84}
M. E. Fisher, J. Stat. Phys. {\bf 34}, 667 (1984).

\bibitem{CK03}
J. Cardy and M. Katori,
J. Phys. A {\bf 36}, 609 (2003).

\bibitem{HF84}
D. A. Huse and M. E. Fisher,
Phys. Rev. B {\bf 29}, 239 (1984).

\bibitem{det}
Such a determinantal expression for non-intersecting
paths was reported in the following mathematical literatures. 
S. Karlin and L. McGregor,
Pacific J. {\bf 9}, 1109, 1141(1959),
B. Lindstr\"om, 
Bull. London Math. Soc. {\bf 5}, {\bf 85} (1973),
I. Gessel and G. Viennot, 
Adv. in Math. {\bf 58}, 300 (1985).
It can be considered as a stochastic version of
Slater determinant of Fermi statistics,
see P.-G. de Gennes, J. Chem. Phys. {\bf 48},
2257 (1968).

\bibitem{KT02a}
M. Katori and H. Tanemura, 
Phys. Rev. E {\bf 66}, 011105 (2002).

\bibitem{PM83}
M. L. Mehta and A. Pandey,
J. Phys. A {\bf 16}, 2655 (1983);
A. Pandey and M. L. Mehta,
Commun. Math. Phys. {\bf 87}, 449 (1983).

\bibitem{NKT03}
T. Nagao, M. Katori and H. Tanemura, 
Phys. Lett. {\bf A307}, 29 (2003).

\bibitem{KKom03}
M. Katori and N. Komatsuda,
Phys. Rev. E {\bf 67}, 051110 (2003).

\bibitem{KGV00}
C. Krattenthaler, A. J. Guttmann, and
X. G. Viennot, 
J. Phys. A {\bf 33}, 8835 (2000).

\bibitem{RY98}
D. Revuz and M. Yor,
{\it Continuous Martingales and Brownian Motion},
3rd ed. (Springer, New York, 1998).

\bibitem{BS96}
A. N. Borodin and P. Salminen,
{\it Handbook of Brownian Motion -- Facts and Formulae},
(Birkh\"auser, Basel, 1996).

\bibitem{Yor92}
M. Yor,
{\it Some Aspects of Brownian Motion, Part I:
Some Special Functionals},
(Birkh\"auser, Basel, 1992).

\bibitem{VZ93}
J. J. M. Verbaarschot and I. Zahed,
Phys. Rev. Lett. {\bf 70}, 3852 (1993).

\bibitem{Ver94}
J. Verbaarschot,
Nucl. Phys. {\bf B426[FS]}, 559 (1994).

\bibitem{JSV96}
A. D. Jackson, M. K. Sener and
J. J. M. Verbaarschot,
Phys. Lett. {\bf B387}, 355 (1996).

\bibitem{SV98}
M. K. Sener and J. J. M. Verbaarschot,
Phys. Rev. Lett. {\bf 81}, 248 (1998).

\bibitem{AZ96}
A. Altland and M. R. Zirnbauer,
Phys. Rev. Lett. {\bf 76}, 3420 (1996).

\bibitem{AZ97}
A. Altland and M. R. Zirnbauer,
Phys. Rev. B {\bf 55}, 1142 (1997).

\bibitem{Nag03}
T. Nagao, 
Nucl. Phys. {\bf B658}, 373 (2003).

\bibitem{FH91}
W. Fulton and J. Harris, 
{\it Representation Theory, 
A First Course},
(Springer, New York, 1991).

\bibitem{Meh91}
M. L. Mehta,
{\it Random Matrices}, 2nd ed.
(Academic, New York, 1991).

\bibitem{KT02b}
M. Katori and H. Tanemura, 
math.PR/0203286.

\bibitem{Dys62}
F. J. Dyson, J. Math. Phys. {\bf 3}, 1191 (1962).

\bibitem{Do84}
J. L. Doob,
{\it Classical Potential Theory and its Probabilistic Counterpart},
(Springer, New York, 1984).

\bibitem{Hua63}
L. Hua,
{\it On the theory of functions of several complex variables. I},
tr. L. Ebner and A. Kor\'ani, 
(American Mathematical Society, Province, RI, 1963).

\bibitem{HC57}
Harish-Chandra,
Am. J. Math. {\bf 79}, 87 (1957).

\bibitem{IZ80}
C. Itzykson and J.-B. Zuber,
J. Math. Phys. {\bf 21}, 411 (1980).

\bibitem{Meh81}
M. L. Mehta, 
Commun. Math. Phys. {\bf 79}, 327 (1981).

\bibitem{Zvo97}
A. Zvonkin,
Matrix integrals and map enumeration:
an accessible introduction,
Math. Comput. Modelling {\bf 26}, 281 (1997);
available from 
http://dept-info.labri.u-bordeaux.fr/~zvonkin/

\bibitem{Oko99}
A. Okounkov,
Int. Math. Res. Not. {\bf 20}, 1043 (2000).

\bibitem{HZ86}
J. Harer and D. Zagier,
Invent. Math. {\bf 85}, 457 (1986).

\bibitem{Bat53}
H. Bateman, {\it Higher Transcendental Functions}, edited by
A. Erd\'elyi,
(McGraw Hill, New York, 1953), Vol.2.

\bibitem{Bat54}
H. Bateman, {\it Tables of Integral Transforms}, edited by 
A. Erd\'elyi, 
(McGraw Hill, New York, 1954), Vol.2.

\bibitem{NF99}
T. Nagao and P. Forrester,
Nucl. Phys. {\bf B563[PM]}, 547 (1999).

\bibitem{Nag01}
T. Nagao,
Nucl. Phys. {\bf B602}, 622 (2001).

\bibitem{KNT03}
M.Katori, T.Nagao and H.Tanemura,
math.PR/0301143;
to be published in
Advanced Studies in Pure Mathematics.

\bibitem{FNH99}
P. J. Forrester, T. Nagao and G. Honner,
Nucl. Phys. {\bf B553[PM]}, 601 (1999).

\bibitem{Red01} S. Redner, 
{\it A Guide to First-Passage Processes}
(Cambridge University Press, Cambridge, 2001).

\bibitem{KR96}
P. L. Krapivsky and S. Redner,
J. Phys. A {\bf 29}, 5347-5357 (1996).

\bibitem{RK99}
S. Redner and P. L. Krapivsky,
Am. J. Phys. {\bf 67}, 1277-1283 (1999).

\bibitem{BB02} A. J. Bray and R. A. Blythe,
Phys. Rev. Lett. {\bf 89}, 150601 (2002).

\bibitem{AJMKR02}
D. ben-Avraham, B. M. Johnson, C. A. Monaco,
P. L. Krapivsky and S. Redner,
J. Phys. A {\bf 36}, 1789 (2003).

\bibitem{Car03}
J. Cardy,
e-print math-ph/0301039.

\bibitem{TK98}
T. Tsuchiya and M. Katori,
J. Phys. Soc. Jpn. {\bf 67}, 1655 (1998).

\bibitem{CC99}
J. Cardy and F. Colaiori,
Phys. Rev. Lett. {\bf 82}, 2232 (1999).

\bibitem{GV02}
A. J. Guttmann and M. V\"oge,
J. Statist. Plann. Inf. {\bf 101},
107 (2002).

\bibitem{IK01}
N. Inui and M. Katori,
J. Phys. Soc. Jpn. {\bf 70}, 1 (2001).

\bibitem{TY03}
H. Tanemura and N. Yoshida,
Probab. Theory and Relat. Fields {\bf 125}, 593 (2003).

\end{thebibliography}

\end{document}